\documentclass[sigconf]{acmart}

\usepackage{multirow}
\usepackage{makecell}
\usepackage{subfigure}
\newcommand{\takeawayBase}[1]{%
    \colorbox{black!5}{%
        \parbox[t]{\dimexpr\linewidth-2\fboxsep-4pt}{
            \vspace*{2pt}%
            \textbf{Takeaway}~\strut#1%
            \vspace*{2pt}%
        }%
    }%
}

\newcommand{\takeawayN}[1]{%
    \noindent %
    \begingroup
    \sbox0{\takeawayBase{#1}}%
    \leavevmode %
    \llap{%
        \hspace*{-0.1pt}%
         \raisebox{\dimexpr -\dp0 - 0.1\baselineskip}{%
            \textcolor{blue!80!black}{%
                \rule[-0.1\baselineskip]{1pt}{\dimexpr \ht0 + \dp0 + 0.2\baselineskip}%
            }%
        }%
        \hspace*{0.1pt}%
    }%
    \usebox{0}%
    \endgroup
    \par%
}

\AtBeginDocument{%
  }

\copyrightyear{2025}
\acmYear{2025}
\setcopyright{cc}
\setcctype{by}
\acmConference[ASIA CCS '25]{ACM Asia Conference on Computer and Communications Security}{August 25--29, 2025}{Hanoi, Vietnam}
\acmBooktitle{ACM Asia Conference on Computer and Communications Security (ASIA CCS '25), August 25--29, 2025, Hanoi, Vietnam}
\begin{document}

\title{When Better Features Mean Greater Risks: The Performance-Privacy Trade-Off in Contrastive Learning}

\author{Ruining Sun}
\email{Seroney_Sun@outlook.com}
\orcid{0009-0000-1328-3773}
\affiliation{%
	\institution{School of Mathematics and Computational Science, \\
    Xiangtan University}
	\city{Xiangtan}
	\state{Hunan}
	\country{China}
}

\author{Hongsheng Hu}
\email{hongsheng.hu@newcastle.edu.au}
\orcid{0000-0003-4455-4227}
\affiliation{%
	\institution{School of Information and Physical Sciences, \\
    University of Newcastle}
	\city{Newcastle}
	\state{NSW}
	\country{Australia}}

\author{Wei Luo}
\email{wei.luo@deakin.edu.au}
\orcid{0000-0002-4711-7543}
\affiliation{%
	\institution{School of Information Technology, \\
    Deakin University}
	\city{Burwood}
	\state{VIC}
	\country{Australia}}

\author{Zhaoxi Zhang}
\email{zhaoxi.zhang-1@student.uts.edu.au}
\orcid{0000-0002-3813-2776}
\affiliation{%
	\institution{School of Computer Science, \\
    University of Technology Sydney}
	\city{Ultimo}
	\state{NSW}
	\country{Australia}}

\author{Yanjun Zhang}
\email{Yanjun.Zhang@uts.edu.au}
\orcid{0000-0001-5611-3483}
\affiliation{%
	\institution{School of Computer Science, \\
    University of Technology Sydney}
	\city{Ultimo}
	\state{NSW}
	\country{Australia}}

\author{Haizhuan Yuan}
\authornotemark[1]
\email{yhz@xtu.edu.cn}
\orcid{0000-0002-2495-0305}
\affiliation{%
	\institution{School of Mathematics and Computational Science, \\
    Xiangtan University}
	\city{Xiangtan}
	\state{Hunan}
	\country{China}}

\author{Leo Yu Zhang}
\authornote{Haizhuan Yuan and Leo Yu Zhang are corresponding authors.}
\email{leo.zhang@griffith.edu.au}
\orcid{0000-0001-9330-2662}
\affiliation{%
	\institution{School of Information and Communication Technology, \\
    Griffith University}
	\city{Southport}
	\state{QLD}
	\country{Australia}}

\renewcommand{\shortauthors}{Sun et al.}

\begin{abstract}
	With the rapid advancement of deep learning technology, pre-trained encoder models have demonstrated exceptional feature extraction capabilities, playing a pivotal role in the research and application of deep learning. However, their widespread use has raised significant concerns about the risk of training data privacy leakage. This paper systematically investigates the privacy threats posed by membership inference attacks (MIAs) targeting encoder models, focusing on contrastive learning frameworks. Through experimental analysis, we reveal the significant impact of model architecture complexity on membership privacy leakage: As more advanced encoder frameworks improve feature‐extraction performance, they simultaneously exacerbate privacy‐leakage risks.
	Furthermore, this paper proposes a novel membership inference attack method based on the $p$-norm of feature vectors, termed the Embedding Lp-Norm Likelihood Attack (LpLA). This method infers membership status, by leveraging the statistical distribution characteristics of the $p$-norm of feature vectors. Experimental results across multiple datasets and model architectures demonstrate that LpLA outperforms existing methods in attack performance and robustness, particularly under limited attack knowledge and query volumes.
	This study not only uncovers the potential risks of privacy leakage in contrastive learning frameworks, but also provides a practical basis for privacy protection research in encoder models. We hope that this work will draw greater attention to the privacy risks associated with self-supervised learning models and shed light on the importance of a balance between model utility and training data privacy.
	Our code is publicly available at: https://github.com/SeroneySun/LpLA\_code.
\end{abstract}





\maketitle

\section{Introduction}
In recent years, self-supervised learning (SSL)~\cite{gui2024survey, uelwer2023survey} has become a powerful model pretraining method, gaining widespread attention due to its efficient training paradigm and transferability.
SSL leverages pretext tasks such as context-based~\cite{mundhenk2018improvements, gidaris2018unsupervised, larsson2017colorization, goyal2019scaling}, masking-based~\cite{zhou2021ibot, bao2021beit, he2022masked, xie2022simmim}, and contrast-based~\cite{wu2018unsupervised, chen2020simple, he2020momentum, grill2020bootstrap} approaches to enable models to train on large-scale unlabeled data and develop powerful encoder mapping inputs to a representation space.
This technology not only addresses the challenges of limited and expensive labeled data but also provides a highly transferable pre-trained encoder, facilitating better performance in downstream tasks such as image classification and object detection.

Among these, contrastive learning has gained prominence due to its stronger transferability and stable, fast convergence during the training process.
However, these models may inadvertently expose private information contained within training data, such as gradient inversion~\cite{zhu2019deep}, inference attack~\cite{shokri2017membership, mehnaz2022your, chaudhari2023snap}, adversarial example~\cite{zhang2022self, Zhang2023Masked}, and data poisoning~\cite{tramer2022truth}.
They expose vulnerabilities in model training and deployment, raising widespread societal concerns and prompting extensive policy discussions.
For example, legal frameworks like the General Data Protection Regulation (GDPR) and the California Consumer Privacy Act (CCPA) impose stringent requirements to protect the collection and usage of user data.
Therefore, a critical challenge in machine learning research is how to effectively protect data privacy while maintaining model performance.

Among various privacy risks studies~\cite{de2020overview}, membership inference attack (MIA)~\cite{hu2022membership, shokri2017membership}, as one of the most prevalent privacy attacks, aims to determine whether a given target sample was part of the training dataset based on model's output (or other information).
The emergence of MIA highlights the vulnerability of data privacy protection in machine learning models, especially when the model is overfitting, as it may leak more crucial information about the training dataset, posing a threat to user's privacy.
This concern is especially pronounced in applications involving sensitive personal information, such as healthcare, finance, or social media.
In these domains, unauthorized usage of user data for commercial profit or malicious exploitation purposes can result in severe harm and losses.

Beyond its role as a privacy attack, MIA has increasingly been adopted as a compliance auditing tool under regulations.
This dual role—as both a privacy threat and a regulatory metric—underscores the critical importance of understanding and mitigating MIA risks, particularly in self-supervised learning (SSL) frameworks where traditional defenses may fall short.
Most existing MIA studies focus on enhancing the efficiency of membership inference in classification tasks~\cite{song2021systematic, li2021membership, carlini2022membership} or other specific tasks~\cite{hayes2017logan, song2020information, gupta2021membership}, whereas how to balance model performance and privacy protection is currently lacking~\cite{liu2022ml, zhang2022evaluating}.
This gap is particularly evident in self-supervised learning models, where traditional MIA strategies are often ineffective due to the models' unsupervised training objectives and frameworks.
These characteristics make understanding the privacy risks in SSL challenging, let alone developing effective mitigation strategies.

\noindent \textbf{Our work:} We take a step further in addressing the above challenges by focusing on the performance-privacy trade-off in contrastive learning.
By constructing an assessment framework for evaluating the utility and privacy of encoders, we analyze the mechanisms of privacy leakage in contrastive learning models and their behavior under membership inference attacks.
Additionally, based on the insights gained from this assessment, we propose a novel attack method that leverages the distribution of feature vectors' $p$-norm, referred to as Embedding Lp-Norm Likelihood Attack (LpLA).

\noindent \textbf{Contributions:} The main contributions are summarized as follows:
\begin{itemize}%[leftmargin=*]
	\item A systematic assessment of the performance-privacy trade-off in contrastive learning. This study reveals, for the first time, that both the different contrastive learning frameworks and the backbone architectures significantly influence the performance and membership privacy leakage of SSL pre-trained encoders. Furthermore, we summarize how different attacks leverage different types of information to perform membership inference attacks.
	
	\item Proposal of a membership inference attack method based on $p$-norm of feature vectors (LpLA). Unlike existing methods that rely on model confidence, loss values, or similarity calculations, LpLA uses the $p$-norm of feature vectors as the attack signal for membership inference. Experiments show that LpLA performs comparably to, or even outperforms, existing attacks across a variety of scenarios, while requiring fewer attack knowledge and query volumes, providing a novel perspective for privacy analysis.
	
	\item Comprehensive experiments for evaluating performance-privacy trade-off and LpLA performance in contrastive learning. Through comparative analysis of multiple attack methods and experimental validation, we offer a comprehensive assessment of the performance-privacy trade-off and LpLA's effectiveness. These findings serve as valuable references for future research on the privacy of contrastive learning models.
	
\end{itemize}

\section{Preliminaries and Related Work}

\subsection{Visual Self-Supervised Representation Learning}
Self-supervised representation learning aims to automatically learn effective and robust feature representations from unlabeled data, providing a solid technical foundation for various downstream tasks in the field of computer vision, such as image classification, object detection, and image segmentation.
To achieve this goal, researchers have proposed a variety of approaches.
In the domain of computer vision, these approaches can generally be categorized into three main types.
Context-based methods~\cite{mundhenk2018improvements, gidaris2018unsupervised, larsson2017colorization, goyal2019scaling} leverage the inherent contextual relationships between data instances (e.g., spatial structures, local-to-global consistency) to construct learning objectives that help models capture more representative features.
Masking-based methods~\cite{zhou2021ibot, bao2021beit, he2022masked, xie2022simmim} typically consist of an encoder-decoder architecture.
The encoder maps partially masked inputs into a low-dimensional feature space to extract critical features, while the decoder attempts to reconstruct the original input based on these features. 

\noindent \textbf{Contrastive-based methods.} Compared with the previous two categories, contrastive learning has gained popularity due to its stronger transferability and stable, fast convergence during the training process.
Unlike context-based and masking-based methods, which train encoders indirectly through complex pretext tasks, contrastive-based self-supervised representation learning directly relies on a simple discrimination task to train the encoder.
Since its introduction in \cite{wu2018unsupervised}, instance discrimination-based contrastive learning methods have quickly become the mainstream approach in this field.
The core mechanism involves generating sample pairs through data augmentation, where each sample is paired with both positive and negative examples.
The model then optimizes a contrastive loss function to bring the feature vectors of positive pairs closer together while pushing apart the feature vectors of negative pairs.
This allows the encoder to learn feature representations capable of distinguishing between similar and dissimilar inputs.
However, early contrastive learning methods were constrained by the simplified design of their training frameworks and the limitations of their data augmentation strategies, which prevented them from fully boosting their efficient feature extraction capabilities.

The capabilities of contrastive learning were fully realized with the introduction of SimCLR and MoCo in \cite{chen2020simple, he2020momentum}.
In \cite{chen2020simple}, positive and negative sample pairs are generated through richer data augmentation, and feature vectors are produced using an encoder and a multi-layer perceptron (MLP) structure.
By treating other samples within the same batch as negatives to compute the contrastive loss, this approach significantly improves the quality of learned representations.
In contrast, \cite{he2020momentum} introduced MoCo for the first time, which proposes a momentum update strategy to maintain a sufficiently large momentum dictionary, ensuring consistency in dictionary representations while constructing sample pairs.
This method not only significantly reduces the computational overhead during training but also further optimizes the effectiveness of feature extraction capabilities.

Subsequently, more sophisticated contrastive learning methods have been proposed.
For instance, SwAV~\cite{caron2020unsupervised} introduces the concept of clustering to replace simple pairwise comparisons, while BYOL~\cite{grill2020bootstrap} and SimSiam~\cite{chen2021exploring} adopt a self-distillation framework to discard negative examples and rely solely on positive examples for optimization.
These innovative approaches have inspired many new research directions in the field of self-supervised representation learning.

\subsection{Membership Inference Attack}
The privacy issues of deep learning models have garnered widespread attention in recent years.
Among privacy risks studies~\cite{de2020overview}, membership inference attack (MIA)~\cite{hu2022membership}, a widely used privacy attack, aims to infer whether a specific data sample was part of a model's training dataset by observing the model's output behavior on the specific sample.
MIA, first introduced by Shokri et al.~\cite{shokri2017membership}, exploits the outputs of machine learning models to infer membership status, raising significant privacy concerns in real-world applications.
Since then, MIA research has rapidly expanded to cover various types of victim models, including regression models~\cite{gupta2021membership}, classification models~\cite{shokri2017membership}, generative models~\cite{hayes2017logan}, and encoder models~\cite{song2020information}.
At the same time, a range of defense methods has been proposed to effectively mitigate MIA attacks while maintaining the utility of the model~\cite{de2020overview, jere2020taxonomy, ma2023loden}.
Now, it's not only a privacy attack on deep learning models but also serves as an audit metric for evaluating the privacy risk of models or algorithms. 

\noindent \textbf{Classic MIAs.} Existing works of MIA mainly focus on classification models, where an adversary leverages various information unintentionally leaked under different settings to perform inference.
Most of these methods follow Shokri using the model's output, which is the most straightforward information, as an attack signal.
Specifically, in a black-box setting~\cite{song2021systematic, li2021membership, carlini2022membership, nasr2019comprehensive}, the adversary often uses the confidence score from a classification model to train a binary classifier.
In contrast, in a white-box setting~\cite{nasr2019comprehensive, melis2019exploiting, rezaei2021difficulty}, the adversary can exploit richer information about the target model (such as gradient information from intermediate layers), combining them into higher-dimensional features, which are then used as an attack signal to infer membership status.

In addition, other types of attack signals have also been explored in previous research for MIA.
For example, \cite{li2022leaks} demonstrated that by calculating entropy or logits from the confidence score, the adversary can use this data as an attack signal and determine a threshold to infer membership status.
In strictly black-box scenarios~\cite{li2021membership}, the adversary can also construct an attack signal using only the label information output by the classification model to complete the inference.
Furthermore, LiRA~\cite{carlini2022membership} stands out for introducing a likelihood ratio statistic as the attack signal, which, in combination with the model's loss distribution on samples, achieves highly accurate membership inference.
LiRA provides a novel perspective for optimizing attacks by focusing on the distribution of statistical metrics.
Beyond classification models, there are also studies on non-classic attack signals in encoder models.
For example, \cite{song2020information} used the similarity between word vectors as an attack signal in text encoders, implementing a highly representative attack method.

\noindent \textbf{MIA against encoder models.} In the field of membership inference attacks targeting visual self-supervised learning, \cite{liu2021encodermi} is the first (and, at the time of writing, the only) study to propose a membership inference attack targeting contrastive learning models.
The method, EncoderMI, leverages the differences in cosine similarity between augmented samples' embedding of member and non-member data to train a binary classifier for distinguishing membership status.
EncoderMI has laid the foundation for subsequent research on membership inference attacks against visual representation learning models.
For example, \cite{gao2023similarity} proposed a strategy to construct attack features based on similarity for the person re-identification task, successfully compromising the privacy of person re-identification encoders under black-box scenarios.
Furthermore, significant progress has been made in membership inference attacks targeting masked pretraining encoders.
For instance, \cite{zhu2024unified} extended the EncoderMI method by computing the similarity between feature vectors of different parts of the same image, achieving more general attack performance.
Similarly, \cite{li2024membership} combined the shadow model technique, leveraging a locally trained shadow decoder to mimic the behavior of the target model and construct a pseudo-loss function, where lower pseudo-loss corresponds to higher membership confidence.

Overall, utilizing the similarity-based attack signal plays an indispensable role in membership inference attacks targeting encoder models.
This is largely due to the common understanding in existing research that, compared to the outputs of classification models, the feature vectors output by encoders often lack explicit and direct semantic information~\cite{liu2021encodermi, gao2023similarity}.
As a result, straightforward strategies that directly use model outputs as attack signal~\cite{hu2022membership, salem2018ml} tend to perform sub-optimally on encoder models.
However, through a series of experiments, we find that under some contrastive learning frameworks, directly using the feature vectors output by encoders can also achieve effective membership inference attacks.
Furthermore, we propose a likelihood estimation attack method based on $p$-norm of feature vectors.
This method demonstrates significant and robust attack performance against encoder models, providing a novel perspective for membership privacy research in encoder models.

\section{Systematically Evaluating Privacy Risks of Encoders}
In this section, we first introduce the framework for evaluating the utility and privacy of encoders.
Next, we specifically introduce several contrastive learning models and existing membership inference attacks, which are used as target encoders and privacy metrics to understand the trade-offs between model utility and privacy risks.
Last, we introduce the experimental settings and summarize the findings from the experimental results.

The systematic evaluation of the privacy risks of encoders aims to answer the two research questions (RQs) as follows:

\begin{itemize}%[leftmargin=*]
	\item \textbf{RQ1}: Do encoders trade off the utility of the model with privacy risks? Put differently, does a higher utility model suffer more from privacy leakage?
	\item \textbf{RQ2}: How do attackers leverage the information produced by the target model to perform membership inference attacks on the target sample?
\end{itemize}

Answering \textbf{RQ1} and \textbf{RQ2} is crucial to understanding the fundamental attack mechanism of membership inference on encoder models.
To answer \textbf{RQ1}, we propose our evaluation mechanism, which consists of two main parts:
The first part measures the feature extraction capability of the encoder model by examining the performance on downstream tasks, which serves as the utility indicator of the encoder;
The second part uses membership inference attack, a widely used privacy attack as the privacy metric, to measure the privacy risks of the encoder.
To answer \textbf{RQ2}, we evaluate benchmarks of membership inference attacks on a series of encoders to obtain the attack performance on them, and we summarize how different attacks leverage different information to perform attacks.

\subsection{Overview of the Evaluation Mechanism}
\begin{figure}[tbp]
	\centering
	\includegraphics[width=0.47\textwidth]{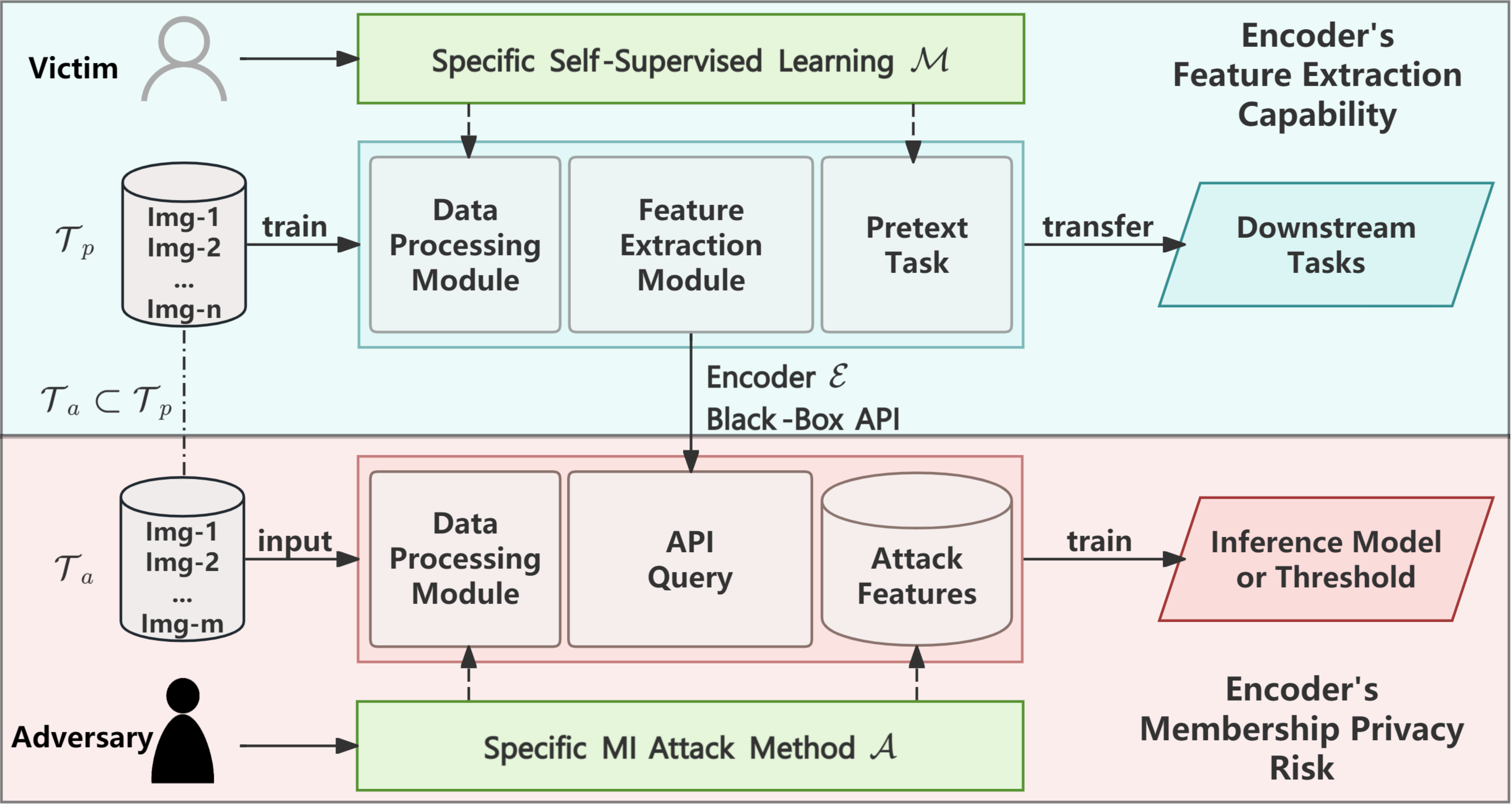}
	\caption{\textbf{An illustration of the evaluation framework. The framework consists of two parts: a utility measurement part for evaluating the model performance and a privacy measurement for evaluating the model's privacy risks.}}
	\label{fig:overview}
\end{figure}

We present the evaluation mechanism in Figure~\ref{fig:overview}.
Specifically, it consists of two parts: the utility measurement part and the privacy measurement part.
Before we detail the two parts, we introduce the threat model in this paper.

\noindent \textbf{Threat model.} We consider that there is a victim who owns a self-supervised learning model, which is referred to target model.
There is also an adversary who has only black-box access to the target model.
That is, the adversary can send query data examples to the target model and receive model outputs.
The adversary's goal is to infer the private information about the training dataset of the target model.
For membership inference attacks in this paper, the attacker is to predict whether a given data sample was in the training dataset or not. 

Given the defined threat model, we now detail the utility measurement and privacy measurement in the evaluation mechanism.

\noindent \textbf{Utility measurement}: To measure the utility of encoder models, we consider that the victim holds an unlabeled private dataset $\mathcal{T}_p$ and uses a specific contrastive learning framework $\mathcal{M}$ for self-supervised training.
After the training process, a pre-trained encoder model $\mathcal{E}$ is obtained, which can be used for transfer learning in downstream tasks.
The encoder $\mathcal{E}(\cdot)$ takes as input an image sample $x$ and outputs its high-dimensional feature vector $v$ in the representation space.
This helps improve the performance of downstream tasks during transfer learning.
We use the performance of $\mathcal{E}$ on downstream tasks as a measure of the encoder model's feature extraction capability.

\noindent \textbf{Privacy measurement}: Following~\cite{zhu2024unified, salem2018ml, liu2022ml}, we consider that the adversary can access a partial training dataset $\mathcal{T}_a \subset \mathcal{T}_p$.
Formally, the adversary's goal is to determine whether a target sample $x$ was used to train the encoder model $\mathcal{E}$ or not, i.e., whether $x_\text{target} \in \mathcal{T}_p$ or not.
In this paper, we consider the standard membership inference game where we pick the target sample from the training dataset with 50\% probability.
To comprehensively measure the privacy risks of the target encoder, the adversary leverages a series of membership inference methods $\mathcal{A}$ that use different attack signals or features to construct the attack model.
We use the performance of the attack model on the target encoder $\mathcal{E}$ as its privacy risk metric.
Intuitively, a better membership inference attack performance means the encoder is less private.

\subsection{Target Contrastive Learning Models}
In this paper, we concentrate on the contrastive learning models of the MoCo series, including the MoCo-v1~\cite{he2020momentum},  MoCo-v2~\cite{chen2020improved}, and MoCo-v3~\cite{chen2021empirical}.
The MoCo series progressively refined its framework to enhance the feature extraction capabilities and demonstrate outstanding performance across multiple visual tasks.
These models have been successfully applied to various encoder architectures, including Convolutional Neural Networks (CNNs)~\cite{he2016deep} and Vision Transformers (ViTs)~\cite{dosovitskiy2020image}.
The core idea of MoCo is to pull together different augmented views of the same image (positive samples) while pushing apart the features of different images stored in the momentum dictionary (negative samples).
We note that there are other contrastive learning models such as SimCLR~\cite{chen2020simple} and BYOL~\cite{grill2020bootstrap} etc.
In this paper, we do not select them because MoCo's progressive refinement facilitates our organized and systematic evaluation of the influence between model complexity and privacy leakage to answer RQ1.
Below, we summarize the major improvements in architectural design, loss optimization objectives, and dictionary maintenance strategies across the different versions of MoCo.

\noindent \textbf{MoCo-v1} is the foundational framework of the MoCo series, which consists of three core components: An image encoder ($f$), a momentum encoder ($f_m$), and a dynamic dictionary ($\Gamma$).
The image encoder $f$ is responsible for generating feature vectors for an augmented input, while the momentum encoder $f_m$, updating at a slower rate using a momentum update strategy, is responsible for generating key vectors for another augmented input.
The dynamic dictionary $\Gamma$ maintains a queue to store key vectors produced by the momentum encoder for inputs from previous mini-batches.

Specifically, given a mini-batch of $N$ inputs, MoCo-v1 generates two augmented versions, $x_\text{q}$ and $x_\text{k}$, for each input $x$.
These augmented inputs are processed by $f$ and $f_m$, to produce a query vector $q_x$ and key vector $k_x$ respectively:
\begin{equation}
	\left\{
	\begin{aligned}
		f(x) &= \text{Pred.}(\mathcal{E}(x)), \\
		q_x &= f(x_\text{q}), \\
		k_x &= f_m(x_\text{k}),
	\end{aligned}
	\right.
\end{equation}
where $\text{Pred.}(\cdot)$ is a linear layer that goes end to end with $\mathcal{E}(\cdot)$ (i.e., serves as the pretext task in Figure~\ref{fig:overview}).
The parameters of image encoder $f$ are optimized using a contrastive loss function (InfoNCE loss~\cite{hadsell2006dimensionality}), while the parameters of momentum encoder $f_m$ are updated using a momentum-based update rule as follows:
\begin{equation}
	\ell(x) = - \log \frac{\exp(\text{sim}(q_x, k_x) / \tau)}{\exp(\text{sim}(q_x, k_x) / \tau) + \sum_{z \in \Gamma} \exp(\text{sim} (q_x, z) / \tau)},  \label{eq:moco-loss}
\end{equation}

\begin{equation}
	\theta_{\mathrm{k}} \leftarrow m \theta_{\mathrm{k}}+(1-m) \theta_{\mathrm{q}}, \label{eq:momentum-update}
\end{equation}
where $\text{sim}(\cdot,\cdot)$ denotes the cosine similarity function, $\tau$ is the temperature parameter, and $m$ controls the update speed of the momentum encoder, which is typically set to 0.999.
MoCo-v1 dynamically updates the dictionary by removing the oldest batch of key vectors and adding newly generated ones after each mini-batch training iteration.
This mechanism ensures continuous optimization by maintaining a diverse and up-to-date set of negative samples in the dictionary.

\noindent \textbf{MoCo-v2} is the second framework of the MoCo series refined based on Moco-v1.
Firstly, it adopted stronger data augmentation strategies, such as multi-scale cropping and color jittering.
Secondly, it replaces the single fully connected layer in $\text{Pred.}$ with a two-layer multilayer perceptron (MLP).
Experimental results demonstrated that this modification produces representations better suited for more transfer learning tasks, enabled MoCo-v2 to achieve performance on multiple unsupervised learning benchmarks closely approached of supervised methods.

\noindent \textbf{MoCo-v3} is the SOTA framework of the MoCo series, which is further refined by introducing significant adjustments in the model framework, loss function, and dictionary maintenance.
Firstly, in terms of the model framework, which is the primary focus of RQ1, the image encoder $f$ now consists of three components: the backbone network $\mathcal{E}$, a projection head $\text{Proj.}$, and a prediction head $\text{Pred.}$.
In contrast, the momentum encoder $f_m$ is composed of only a backbone network $\mathcal{E}$ and a projection head $\text{Pred.}$, where $\text{Pred.}$ and $\text{Proj.}$ are all constructed using fully connected layers, normalization layers, and stacked ReLU activation functions,
\begin{equation}
	\left\{
	\begin{aligned}
		f(x) &= \text{Proj.}(\text{Pred.}(\mathcal{E}(x))), \\
		f_m(x) &= \text{Pred.}(\mathcal{E}(x)).
	\end{aligned}
	\right.
\end{equation}

Additionally, there are also changes in the loss function and dictionary maintenance strategy.
Instead of maintaining a large dictionary $\Gamma$, MoCo-v3 uses all non-self samples within the mini-batch as negative samples for loss computing.
Under this configuration, given two augmented versions $x_1$ and $x_2$ of each input $x$, there would be two pairs of output from $f$ and $f_m$ respectively, denoted by query vectors $q_1$, $q_2$ and key vectors $k_1$, $k_2$ respectively:
\begin{equation}
	q_i = f(x_i),\text{ } k_i = f_m(x_i), \text{ } i=1,2.
\end{equation}

In the framework of MoCo-v3, the contrastive loss is then computed as follows:
\begin{equation}
	\left\{
	\begin{aligned}
		\ell(x_1) &= - \log \frac{\exp(\text{sim}(q_1, k_2^+) / \tau)}{\exp(\text{sim}(q_1, k_2^+)) / \tau) + \sum_{z \in x_2^-} \exp(\text{sim}(q_1, z) / \tau)}, \\
		\ell(x_2) &= - \log \frac{\exp(\text{sim}(q_2, k_1^+) / \tau)}{\exp(\text{sim}(q_2, k_1^+)) / \tau) + \sum_{z \in x_1^-} \exp(\text{sim}(q_2, z) / \tau)}, \\
		\mathcal{L}(x) &= \ell(x_1) + \ell(x_2),
	\end{aligned}
	\right.
\end{equation}
where $x^-$ denotes non-self samples within the same mini-batch, and $\mathcal{L}(x)$ represents the final optimization objective used to update $f$.
The momentum encoder $f_m$ is still updated in the same manner using equation \eqref{eq:momentum-update} (obviously, updating $\mathcal{E}$ and $\text{Pred.}$ only).
Through these adjustments, MoCo-v3 achieves much better performance on many downstream tasks and is more suitable for the ViT type backbones.

\noindent \textbf{Why MoCo series are selected.} The MoCo series have progressively evolved from MoCo-v1 to v3. 
Innovations such as momentum updates, dynamic dictionaries, the upgrade of prediction head modules, and the introduction of projection head modules have not only significantly improved the performance of pre-trained encoders but also expanded their applicability to more types of backbone.

Meanwhile, they facilitate our organized and systematic evaluation of the influence between model complexity (contrastive learning framework and backbone architecture) and privacy leakage in encoder models, helping us answer \textbf{RQ1}.

\subsection{Benchmarks of Membership Inference Attacks}
In this section, we present an overview of a suite of membership inference attacks proposed in prior studies~\cite{liu2021encodermi,gao2023similarity,liu2022ml}, including EncoderMI~\cite{liu2021encodermi}, SD-MI~\cite{gao2023similarity}, and Feature-based MI~\cite{liu2022ml}.
These attacks serve as the privacy risk evaluation foundation for assessing the target models' privacy leakage.

\noindent \textbf{Feature-based MI.} In existing works of membership inference attacks on contrastive learning models, one of the most common attack methods is feature-based MI (referred to as Fe-MI).
When targeting classification models, attackers can use the confidence scores or class labels directly as the attack signal for membership inference~\cite{shokri2017membership}, and there are numerous studies~\cite{salem2018ml, liu2022ml} having widely demonstrated its effectiveness against classification models.
However, previous research~\cite{liu2021encodermi, gao2023similarity} suggests that the feature vectors output by encoder models primarily serve as representations and lack specific semantic information.
Consequently, directly using these feature vectors for binary classifier training could not achieve effective attack performance as in classification models.

To comprehensively explore the potential membership inference attack signal and privacy risks faced by encoder models, we consider Fe-MI as a baseline in the evaluation experiments.
Counter-intuitively, as we will show in Section~\ref{subsec:part1ExpResults}, our experimental results demonstrate that this method can also achieve successful inference attacks against encoder models in some attack scenarios.

The adversary directly collects the feature vectors $v$ through the target encoder's API to form an attack dataset, and uses a simple neural network as the attack model (denoted by $Attacker$), which is detailed as follows:
\begin{equation}
	\mathcal{I}(x)= \text{Attacker}(v_x).
	\label{eq:simple-attacker-fe}
\end{equation}

\noindent \textbf{Similarity-based MI.} In addition to feature-based MI, there are some MI attack methods specifically designed for encoder models, which typically rely on similarity between feature vectors to construct the attack model.
In this paper, we use two representative attack methods as baselines for evaluation: EncoderMI~\cite{liu2021encodermi} and SD-MI~\cite{gao2023similarity}.
We will refer to them as similarity-based MI for simplification.

\noindent \textbf{EncoderMI} is the first membership inference attack method designed for contrastive learning pre-trained encoders.
It leverages the inherent characteristics of contrastive learning optimization objectives, wherein contrastive learning tends to generate similar feature vectors for different augmented views of one input.

To perform membership inference, EncoderMI first generates $n$ augmented views of a given image $x$ using the same data augmentation strategies as the target encoder's training.
These augmented views are denoted as \{$x_1, x_2, \dots, x_\text{n}$\}.
Subsequently, these views are used to query the model API and produce the corresponding feature vectors $V_x = \{v_x^1, v_x^2, \dots, v_x^\text{n}\}$.
As for a specific target sample $x$, EncoderMI computes an attack feature based on the pairwise similarities of the vectors in $V_x$ as follows:
\begin{equation}
	Sig(x) = \{\text{sim}(v_x^i, v_x^j) \,|\, \text{ }i,j=1,2,\dots,\text{n}, \text{ }j > i \},
\end{equation}
where $\text{sim}(\cdot,\cdot)$ represents the cosine similarity.
EncoderMI then sorts the resulting $\text{n} \cdot (\text{n} - 1) / 2$ similarity scores, and uses these sorted $Sig(x)$ as the attack signal to train an inference model, which is usually a simple binary classifier as follows:
\begin{equation}
	\mathcal{I}(x)= \text{Attacker}(Sig(x)). 
	\label{eq:simple-attacker-en}
\end{equation}

\noindent \textbf{SD-MI} is another similarity distribution-based membership inference attack against Re-ID models (Person Re-Identification), which is a type of representation learning, whose training objective is similar to contrastive learning.
Re-ID targets optimizing an encoder to map the features of paired (or grouped) image samples as closely as possible while ensuring that the feature mappings of unpaired (or different groups of) image samples remain distinct.

Specifically, SD-MI first randomly selects anchor images $X_{\text{anchors}}$ to construct a reference dataset, and query $\mathcal{E}(\cdot)$ with anchors to create an ordered set of feature vectors $V_\text{anchor} = \{v_\text{anc}^1, v_\text{anc}^2, \dots, v_\text{anc}^n\}$.
Subsequently, the similarity between target image features $v_x$ and these ordered anchor features $V_\text{anchor}$ is calculated as a similarity distribution of the target image, which serves as the attack signal of SD-MI.
The formulation is as follows:
\begin{equation}
	Sig(x) = [\textnormal{dist}(v_x, v_\text{anc}^1), \textnormal{dist}(v_x, v_\text{anc}^2), \dots, \textnormal{dist}(v_x, v_\text{anc}^n)],
\end{equation}%
where $\textnormal{dist}(\cdot, \cdot)$ represents the Euclidean distance between two feature vectors.
SD-MI then feeds this similarity distribution $Sig(x)$ into a binary attack classifier for membership inference.

To further enhance the attack performance, SD-MI introduces a mechanism named \textit{Anchor Selector}, which assigns weights to the similarity distribution $Sig(x)$ based on the target sample's feature vector $v_x$.
By assigning different weights, it enhances the privacy leakage caused by $Sig(x)$ and enables a significant attack improvement compared to the EncoderMI.
The complete SD-MI consists of two modules as follows:
\begin{equation}
	\mathcal{I}(x)= \text{Attacker}(Sig(x) \odot \text{Selector}(v_x)),
\end{equation}
where $\odot$ denotes the Hadamard product.

\noindent \textbf{Summary of baseline attacks.} The three baseline attack methods are selected for a comprehensive and rigorous evaluation of the privacy risks of encoder models.
They respectively represent three different attack mechanisms where each of them considers a different type of attack signal for leaking the information of the training data:
(1) the encoder model's output vector $v$ itself; (2) the similarity between outputs of the same sample; (3) the similarity of the target sample with auxiliary datasets.
Using such baseline attacks for evaluation helps us to answer \textbf{RQ2}: they enable a comprehensive analysis and comparison of the performance across different attack signals for membership inference against encoder models.

\subsection{Experimental Setting}
\noindent \textbf{Datasets.} We select three commonly used datasets in membership inference attack research for experimental evaluation: CIFAR-10, CIFAR-100, and Tiny-ImageNet.
These datasets encompass different categories and complexities, effectively satisfying the requirements of our experimental evaluation.
As Table~\ref{tab:default_details} shows, we assume the target model is pre-trained with a privacy dataset (20k samples), and tested with a test dataset (10k samples).
On the other hand, an adversary uses an attack dataset composed of 2k samples from the privacy dataset and 2k from the test dataset separately.
{And there is another inference test dataset that is used to evaluate the attack performance, composed of 8k samples from the privacy dataset and 8k from the test dataset disjointed with the attack dataset.}
\begin{itemize}%[leftmargin=*]
	\item CIFAR~\cite{krizhevsky2009learning}: The CIFAR-10 dataset contains 60,000 color images divided into 10 categories, with each image having a resolution of 32×32×3. Similarly, the CIFAR-100 dataset contains 60,000 color images but with an extended 100 categories, and the images maintain the same resolution of 32×32×3.
	\item Tiny-Imagenet~\cite{Le2015TinyIV}: This dataset consists of 200 categories, with a total of 100,000 training images and 10,000 test images. Each image has a resolution of 64×64×3.
\end{itemize}

\noindent \textbf{Target models.} Concretely, we consider four widely used backbone networks as the feature extractors, i.e., ResNet-18, ResNet-50~\cite{he2016deep}, ViT-Small, and ViT-Base~\cite{dosovitskiy2020image} in three versions of the MoCo model.
Following the setting in~\cite{Susmelj_Lightly}, we set the size of the momentum dictionary to 16,384 in MoCo-v1 and MoCO-v2, which differs from the official setting of 65,536.
This choice is made because a dictionary size that is as large as possible while still smaller than the total training sample size is more conducive to achieving better training performance.
Furthermore, following similar settings in~\cite{liu2021encodermi, zhu2024unified}, all target models were trained for 2,000 epochs.

\noindent \textbf{Attack model.} We follow the existing works of EncoderMI~\cite{liu2021encodermi} and Fe-MI to use a simple three-layer multilayer perceptron structure with ReLU activation functions as the attack model, training with 200 epochs.
For SD-MI~\cite{gao2023similarity}, we adopt the official model design and attack settings, which include an \textit{Anchor Selector} composed of two linear layers with ReLU activation functions, and an attack model consisting of five linear layers with Tanh activation functions.

\noindent \textbf{Metrics.} To measure the utility of the target model, we follow the contrastive learning works and use the $K$-NN clustering algorithm~\cite{wu2018unsupervised} to classify the testing samples based on their representative vectors.
A higher accuracy of the $K$-NN clustering algorithm means a better feature extraction capability of the encoder.
To measure the privacy risks of the encoder model, we use the attack performance of the membership inference attack as a metric.
To quantify the attack performance, we use four widely used metrics of accuracy, precision, recall, and TPR@0.1\%FPR, as the membership inference attack task is essentially a binary classification problem.

\noindent \textbf{Encoder training.} We use the official public code for training MoCo-v1, MoCo-v2\footnote{MoCo-v1\&v2: \url{https://github.com/facebookresearch/moco}}, and MoCo-v3\footnote{MoCo-v3: \url{https://github.com/facebookresearch/moco-v3}}, $K$-NN clustering\footnote{InstDisc: \url{https://github.com/zhirongw/lemniscate.pytorch}}, and membership inference attack\footnote{SD-MI: \url{https://github.com/Vill-Lab/2023-AAAI-SDMIA}}.
Additional training configurations are presented in Table \ref{tab:default_details}.
\begin{table}[t]
	\centering
	\caption{Default implementation details in the experiments.}
	\resizebox{0.48\textwidth}{!}{
		\begin{tabular}{lccccc}
			\toprule
			\textbf{Model}  &\textbf{Train-Set}  &\textbf{Test-Set}  &\textbf{Batch-Size}  &\textbf{Epoch}  &\textbf{Other}\\
			\midrule
			MoCo-v1\&2 & $\mathcal{T}_p$ (20k samples) & (None)     & 256     & 2k     & 16,384 for $\Gamma$  \\
			MoCo-v3    & $\mathcal{T}_p$ (20k samples) & (None)     & 512     & 2k     & (None)  \\
			\midrule
			$K$-NN test  & $\mathcal{T}_p$ (20k samples) & $\mathcal{T}_{test}$ (10k samples) & (None)  & (None) & $k$=20  \\
			\midrule
			EncoderMI  &$(2k \in \mathcal{T}_p) \cup (2k \in \mathcal{T}_{test})$ & 
			$(8k \in \mathcal{T}_p) \cup (8k \in \mathcal{T}_{test})$ & 
			128    & 200    & 10 augmentation  \\
			SD-MI      &$(2k \in \mathcal{T}_p) \cup (2k \in \mathcal{T}_{test})$ & 
			$(8k \in \mathcal{T}_p) \cup (8k \in \mathcal{T}_{test})$ & 
			128    & 200    & 2k anchors  \\
			Fe-MI      &$(2k \in \mathcal{T}_p) \cup (2k \in \mathcal{T}_{test})$ & 
			$(8k \in \mathcal{T}_p) \cup (8k \in \mathcal{T}_{test})$ & 
			128    & 500    & (None)  \\
			\bottomrule
	\end{tabular}}
	\label{tab:default_details}
\end{table}

\subsection{Experiment Results}
\label{subsec:part1ExpResults}
We first address the \textbf{RQ1:} ``Do encoders trade off the utility of the model with privacy risks? Put differently, does a higher utility model suffer more from privacy leakage?''
To answer this question, we present the $K$-NN classification performance (i.e., encoder's utility) and membership inference attack performance (i.e., encoder's privacy) against the encoders trained on CIFAR-100 in Table~\ref{tab:Result_overview}.

\begin{table*}[tbp]
	\centering
	\caption{The $K$-NN classification accuracy and membership inference attack performance against different encoder models using a different feature extractor.}
	\resizebox{1.0\textwidth}{!}{
		\begin{tabular}{ccccccccccccccc}
			\toprule
			\multirow{2}[4]{*}{\textbf{Backbone}} & \multirow{2}[4]{*}{\textbf{Model}} & \multirow{2}[4]{*}{\textbf{$K$-NN Accuracy}} & \multicolumn{4}{c}{\textbf{Encoder-MI}} & 
			\multicolumn{4}{c}{\textbf{SD-MI}} & \multicolumn{4}{c}{\textbf{Fe-MI}} \\
			\cmidrule(lr){4-7}  \cmidrule(lr){8-11} \cmidrule(lr){12-15} &    & \multicolumn{1}{c}{} & \textbf{Accuracy} & \textbf{Precision} & \textbf{Recall} & \textbf{TPR@0.1\%FPR} & \textbf{Accuracy} & \textbf{Precision} & \textbf{Recall} & \textbf{TPR@0.1\%FPR} & \textbf{Accuracy} & \textbf{Precision} & \textbf{Recall} & \textbf{TPR@0.1\%FPR} \\
			\midrule
			\multirow{3}[2]{*}{\makecell{ResNet-18}} & MoCo-v1 & 0.427 & 0.500 & 0.500 & 0.522 & 0.100 & 0.510 & 0.513 & 0.407 & 0.105 & 0.505 & 0.506 & 0.373 & 0.103 \\
			& MoCo-v2 & 0.523  & 0.592 & 0.595 & 0.576 & 0.147 & 0.674 & 0.685 & 0.644 & 0.217 & 0.618 & 0.608 & 0.660 & 0.155 \\
			& MoCo-v3 & 0.521  & 0.747 & 0.744 & 0.755 & 0.291 & 0.716 & 0.755 & 0.640 & 0.309 & 0.755 & 0.770 & 0.728 & 0.336 \\
			\midrule
			\multirow{3}[2]{*}{\makecell{ViT-Small}} & MoCo-v1 & 0.416 & 0.518 & 0.520 & 0.453 & 0.109 & 0.494 & 0.494 & 0.510 & 0.098 & 0.508 & 0.508 & 0.810 & 0.104 \\
			& MoCo-v2 & 0.509  & 0.515 & 0.517 & 0.432 & 0.107 & 0.514 & 0.513 & 0.551 & 0.105 & 0.509 & 0.509 & 0.503 & 0.104 \\
			& MoCo-v3 & 0.562  & 0.622 & 0.606 & 0.699 & 0.154 & 0.658 & 0.616 & 0.841 & 0.161 & 0.788 & 0.792 & 0.782 & 0.382 \\
			\bottomrule
	\end{tabular}}
	\label{tab:Result_overview}
\end{table*}

Table~\ref{tab:Result_overview} indicates that as the encoder model becomes progressively more sophisticated, the classification accuracy of the $K$-NN algorithm on the representative vectors of the testing samples improves correspondingly.
This reflects an enhancement in the feature extraction capability of the encoders, i.e., the encoder's utility is better from MoCo-v1 to MoCo-v3.
However, alongside the increase in model utility, a notable rise in membership inference accuracy, precision, recall and TPR@0.1\%FPR can also be observed.
For example, when the target encoder is trained using the simplest MoCo-v1 framework, the accuracy of all attack methods is close to 50\% (i.e., random guess), indicating that the adversary cannot infer the membership information.
As the MoCo framework is progressively upgraded, the accuracy of all membership inference methods increases against target models trained with MoCo-v2 and MoCo-v3.
This reveals that while more complex contrastive learning frameworks enhance feature extraction capabilities, they also lead to higher risks of privacy leakage.

\begin{figure}[tbp]
	\centering
	\subfigure[More complex: ResNet]{\includegraphics[width=4.0cm]{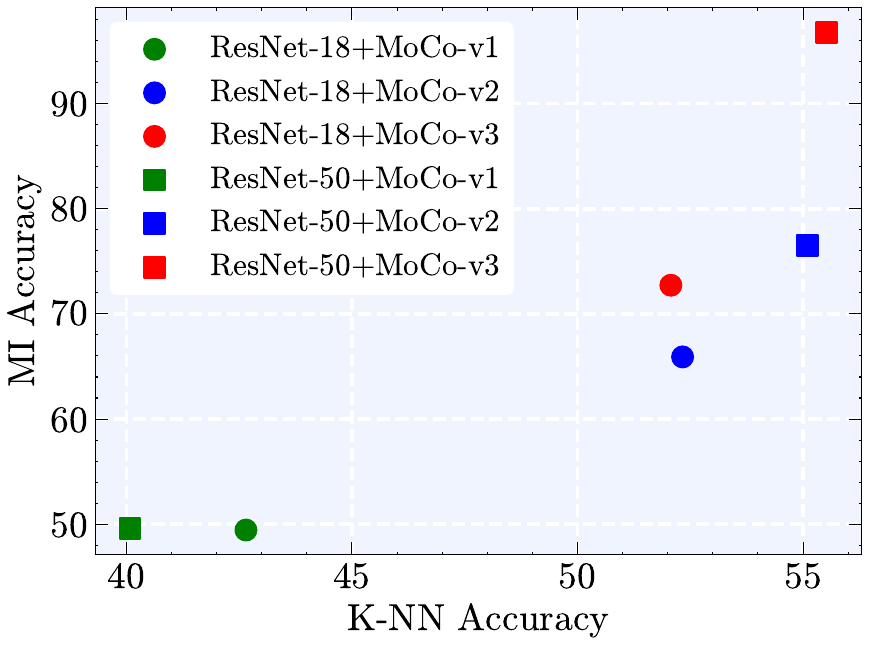}}
	\subfigure[More complex: ViT]{\includegraphics[width=4.0cm]{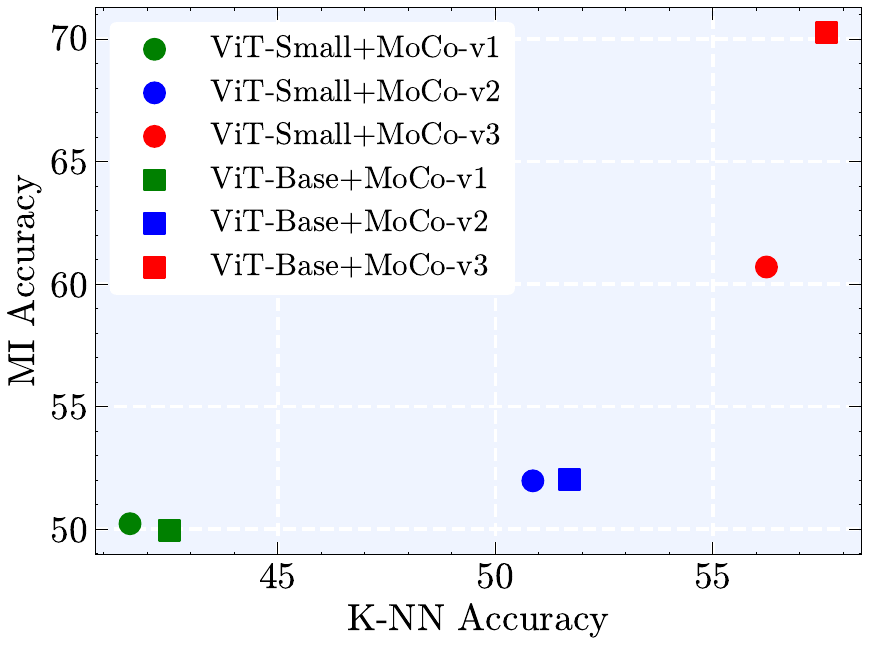}}
	\vspace{-0.2cm}
	\caption{An illustration of the performance-privacy trade-off when the encoder model becomes progressively more sophisticated. Not only a complex contrastive learning framework, but an advanced backbone leads a higher utility and privacy risks.}
	\label{fig:Result_model_desine}
\end{figure}

To further support the finding that the encoder models trade their utility for privacy risks, we conduct the experiments on a series of backbone feature extractors, i.e., the ResNet family of ResNet18 and ResNet50 and ViT family of ViT-Small and ViT-Base, across MoCo-v1, MoCo-v2, and MoCo-v3.
The experimental results are provided in Figure~\ref{fig:Result_model_desine}.
In each plot in Figure~\ref{fig:Result_model_desine}, the x-axis represents the model utility and the y-axis represents the model's privacy risks.
A data point closer to the bottom right indicates a model with both high utility and privacy preservation.
As we can see, the more advanced contrastive learning framework can indeed improve the feature extraction capabilities of the encoders, while the risk of privacy leakage of the model is also increased.
In addition, an interesting finding is that as the complexity of backbone architecture increases (e.g., replacing ResNet18 to ResNet50, or replacing ViT-Small to ViT-Base), the $K$-NN classification accuracy improves, while the membership inference accuracy exhibits a simultaneous increase.
This further suggests that the encoder model inherently trades utility for heightened privacy risks.

\takeawayN{\textbf{1:} In contrastive learning, encoder models trade off the utility with privacy risks. An encoder model having a higher utility usually leads to higher privacy leakage risks.}

\begin{figure}[tbp]
	\centering
	\subfigure[ResNet-18 as the feature extractor]{\includegraphics[width=4.0cm]{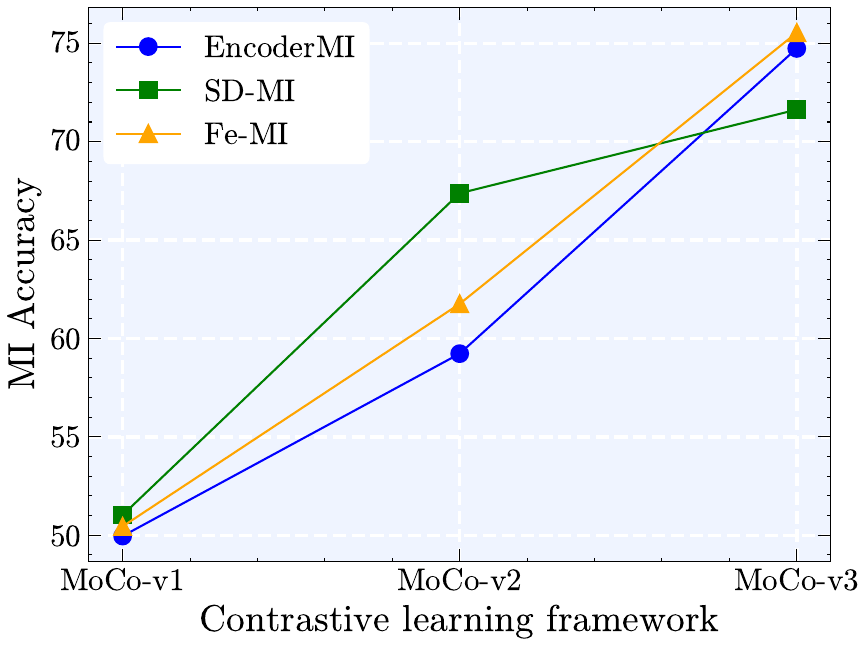}}
	\subfigure[ResNet-50 as the feature extractor]{\includegraphics[width=4.0cm]{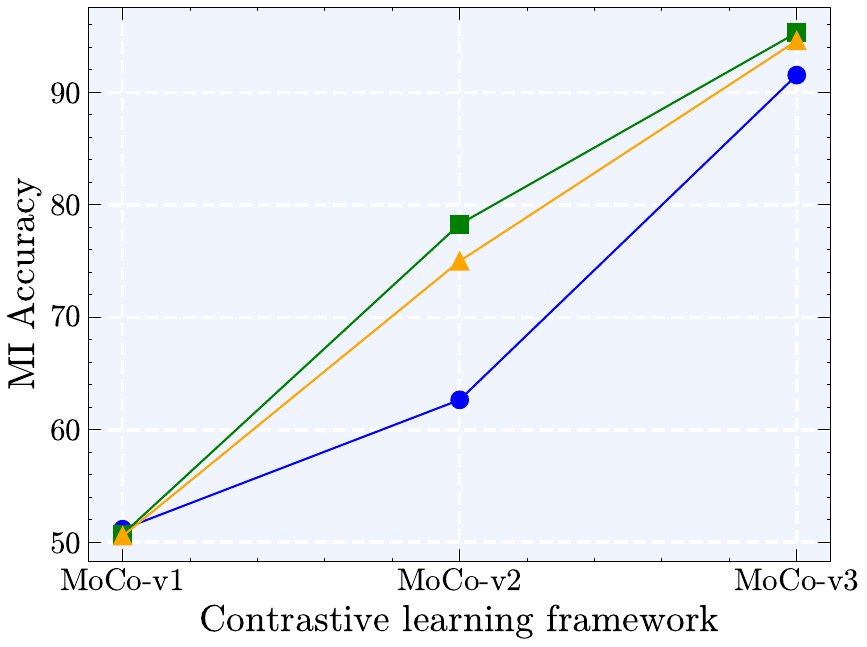}}
	\subfigure[ViT-Small as the feature extractor]{\includegraphics[width=4.0cm]{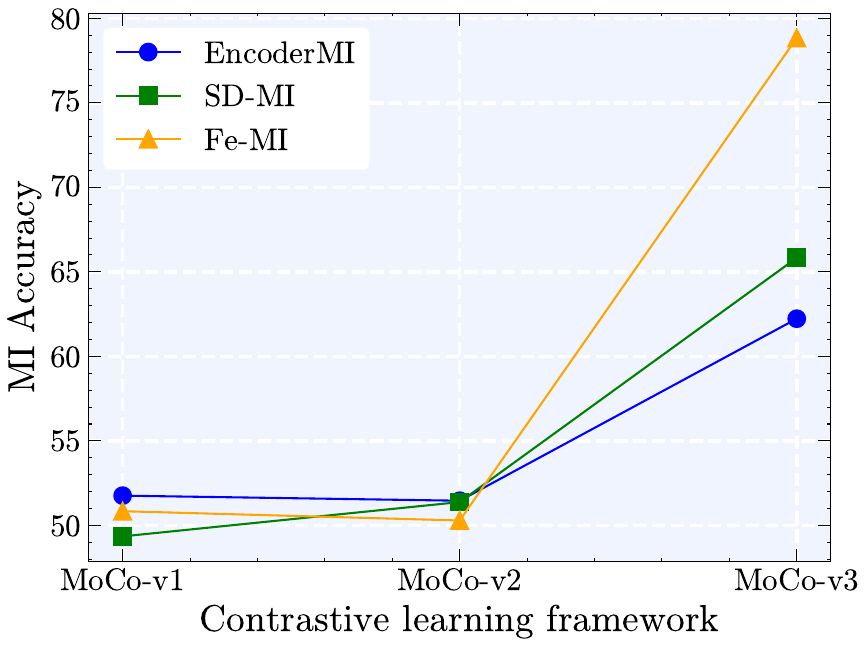}}
	\subfigure[ViT-Base as the feature extractor]{\includegraphics[width=4.0cm]{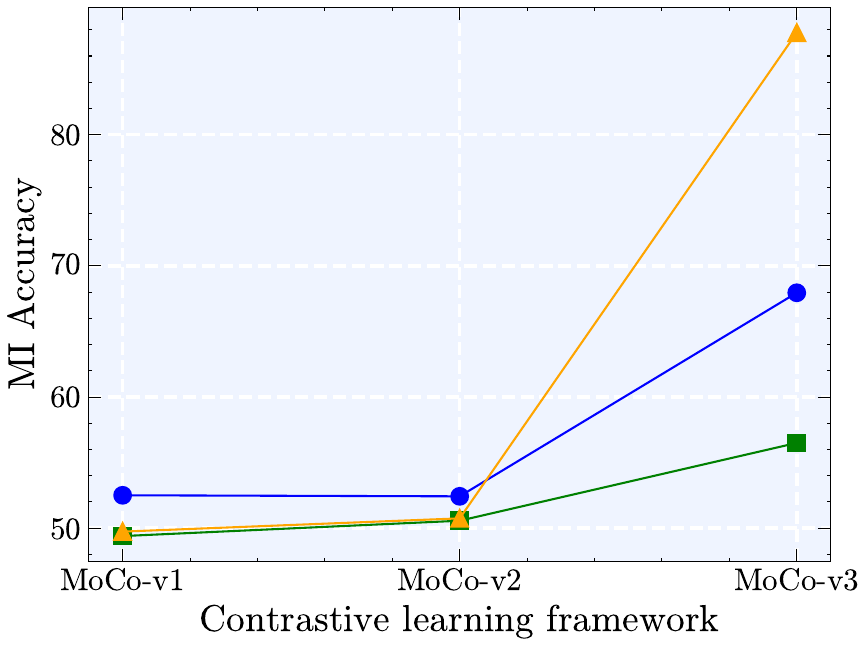}}
	\vspace{-0.2cm}
	\caption{An illustration of the three attacks on different feature extractors trained with MoCo frameworks. Fe-MI achieves a notable performance, indicating that the feature vector as the membership signal itself can be leveraged for MIA directly.}
	\label{fig:Result_signals}
\end{figure}

We now address the \textbf{RQ2:} ``How do attackers leverage the information produced by the target model to perform membership inference attacks on the target sample?''
To answer this question, we compare the performance of different attacks, where each uses different information as the membership signal.
We present the experimental results of the three attacks on different feature extractors trained with MoCo-v1, MoCo-v2, and MoCo-v3 in Figure~\ref{fig:Result_signals}.

From Figure~\ref{fig:Result_signals}, we find that Fe-MI not only achieves highly competitive attack performance under several target training settings but even performs on par with (and in some cases exceeded) the other two similarity-based methods under the MoCo-v3 framework.
Note that Fe-MI directly leverages the feature vector of the target sample for membership inference, while EncoderMI and SD-MI convert the feature vector into the similarity information for the attack.
The experimental results suggest that the feature vector of the target sample in the representative space contains abundant information for facilitating the membership inference attacks.
Unlike similarity-based membership signals using the feature vector as the ``side'' information, the feature vector as the membership signal itself can directly be leveraged by neural networks for conducting the attack. 

However, the Fe-MI method still requires training a deep neural network as the binary attack classifier, which can be computationally expensive.
This limitation motivates us to design a new membership inference attack directly using the feature vector while outperforming the Fe-MI method.

\takeawayN{\textbf{2:} The feature vector of the target sample and similarity scores based on the feature vector can be leveraged by the adversary for a successful membership inference.
	Compared to similarity scores-based membership inference, directly using the feature vector as the membership inference signal shows a consistent competent attack performance in various settings.}

\section{Our Attack}
In this section, we introduce a new membership inference attack, which directly leverages the feature vector of a target sample to inferring its membership status.
Before we go to the details of our proposed attack, we ask the following research questions (RQs): 
\begin{itemize}%[leftmargin=*]
	\item \textbf{RQ3}: How to design a membership inference attack that can directly use the feature vector of the sample as the membership signal but without training a binary attack classifier? 
	\item \textbf{RQ4}: What are the advantages of the proposed attack compared to existing methods?
\end{itemize}
Answering these two questions is crucial for understanding our attack mechanism and can shed light on how easily encoder models may leak their private information about their training data.

\subsection{Motivation of the Attack}
We first address \textbf{RQ3}: ``How to design a membership inference attack that can directly use the feature vector of the sample as the membership signal but without training a binary attack classifier?''
To answer this question, we first introduce the intuition of the attack and then present the design details of the attack method.

\begin{figure}[tbp]
	\centering
	\includegraphics[width=0.37\textwidth]{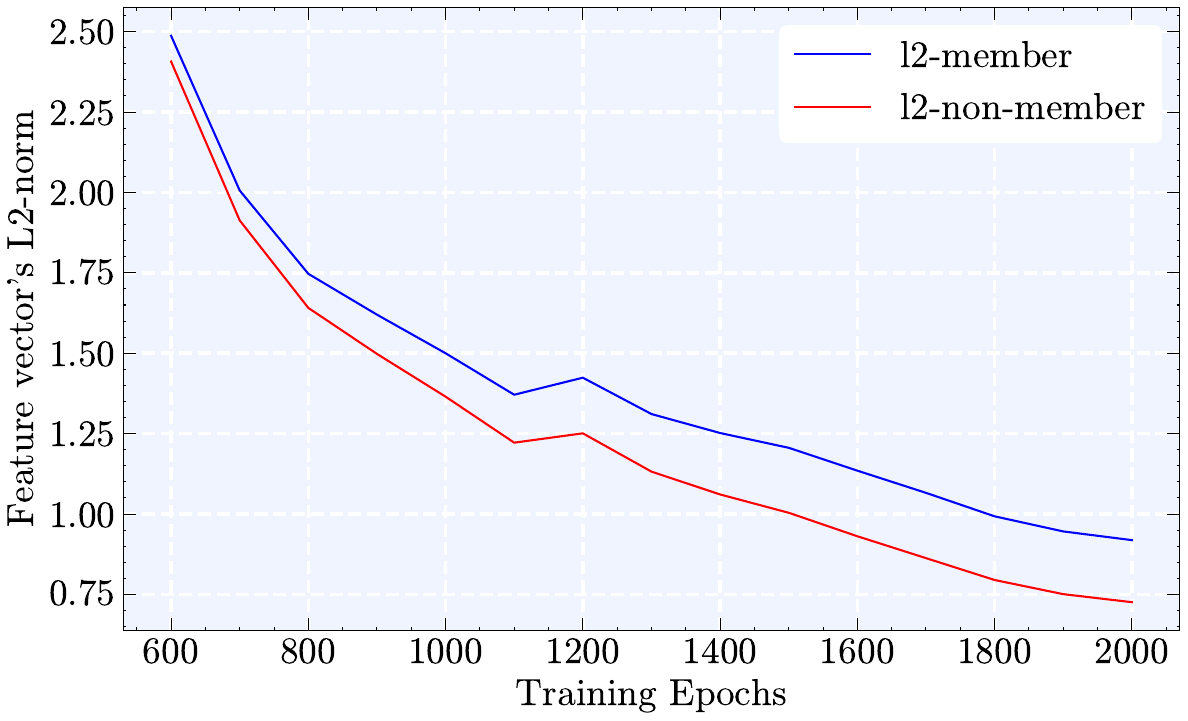}
	\caption{An illustration of the mean L2-norm value of the feature vectors in members and non-members in ResNet-50 trained by MoCo-v2. An obvious distinction of L2-norm values between the two classes of samples can be observed in the training process.}
	\label{fig:local_traing_process}
\end{figure}

\begin{figure*}[tbp]
	\centering 
	\subfigure[L0-Norm]{\includegraphics[width=4.3cm]{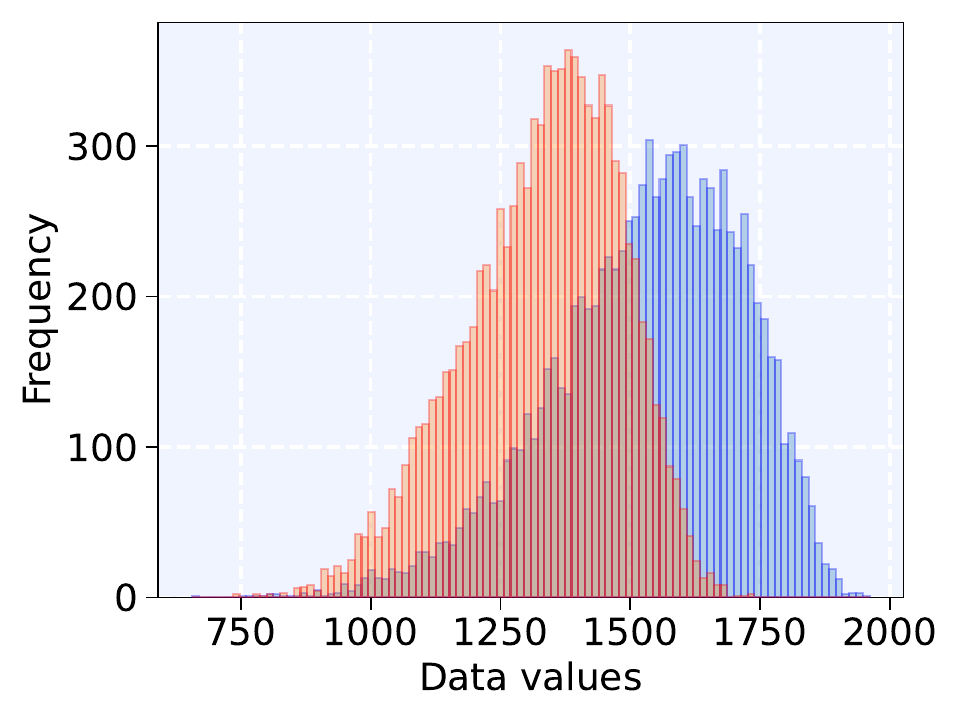}\label{fig:official_v2_hist_a}}
	\subfigure[L1-Norm]{\includegraphics[width=4.3cm]{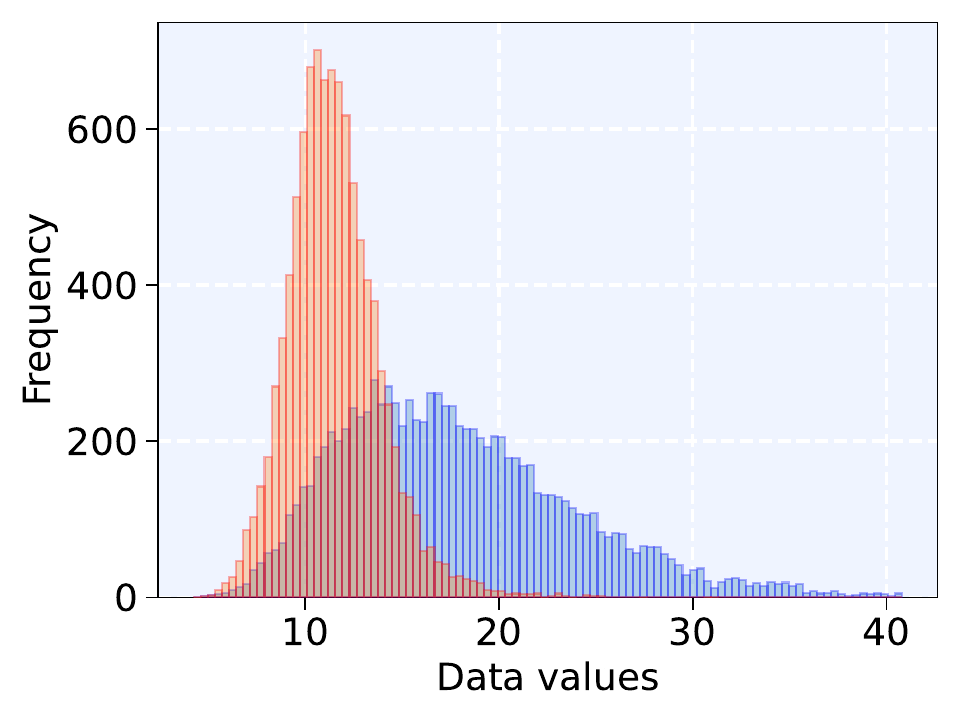}\label{fig:official_v2_hist_b}}
	\subfigure[L2-Norm]{\includegraphics[width=4.3cm]{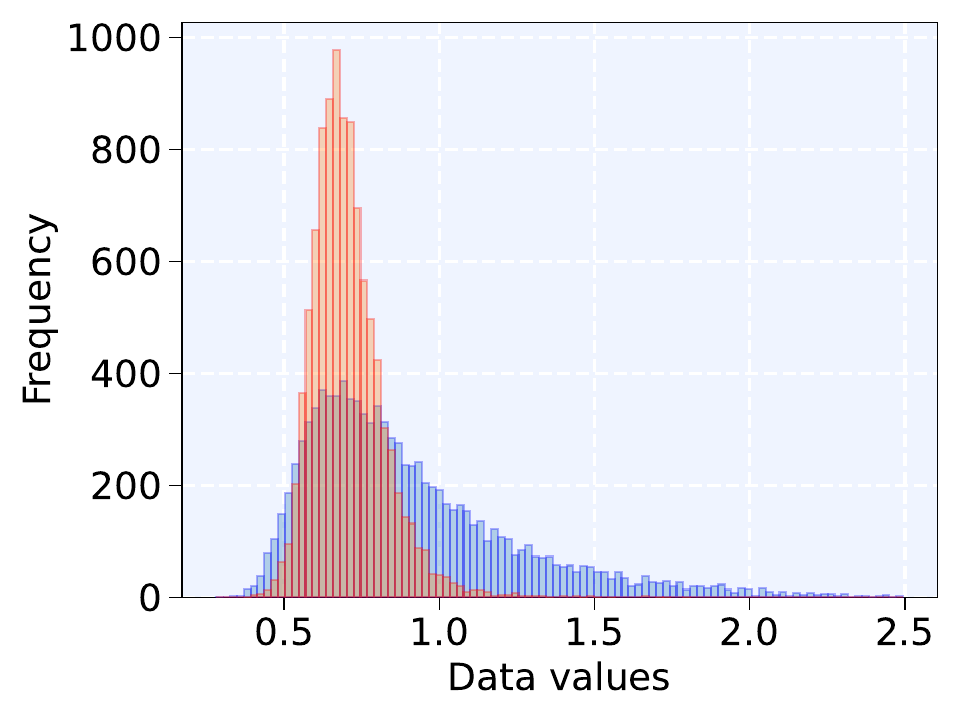}\label{fig:official_v2_hist_c}}
	\subfigure[L3-Norm]{\includegraphics[width=4.3cm]{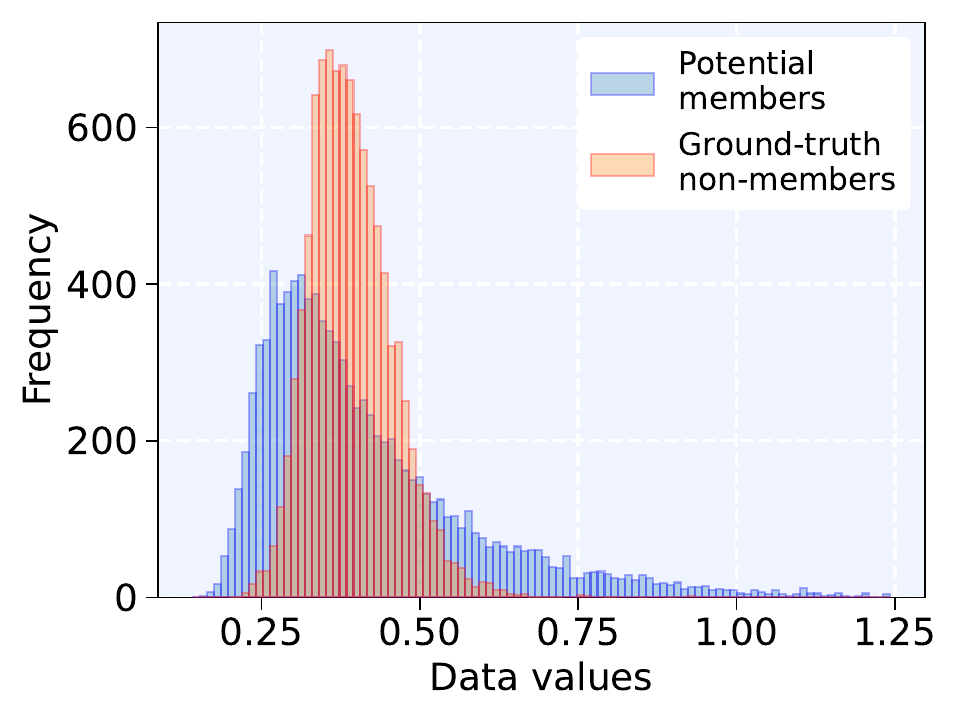}\label{fig:official_v2_hist_d}}
	\vspace{-0.3cm}
	\caption{An illustration of the $p$-norm of feature vectors produced by official pre-trained ResNet-50 with MoCo-v2 framework. There is a significant difference in the feature vector magnitude distributions between the two classes, wherein the behavior aligns closely with the observation highlighted in Figure~\ref{fig:local_traing_process}.}
	\label{fig:official_v2_hist}
\end{figure*}

As we observed in \textbf{RQ2}, the feature vector of the target sample contains abundant information for membership inference.
We take a further look at this vector by examining its magnitude.
Specifically, we calculate the L2-norm value of the vectors of members and non-members throughout the training epochs using the ResNet50 feature extractor in the MoCo-v2 framework.
As shown in Figure~\ref{fig:local_traing_process}, we notice a notable difference in the mean L2-norms of feature vectors between member and non-member samples.
As the training progressed, the L2-norm of feature vectors of non-member samples consistently surpassed those of members.

To further support this observation, we conduct experiments on the official pre-trained ResNet-50 encoder model released by MoCo team\footnote{moco-v2-800ep-pretrain.pth.tar: \url{https://dl.fbaipublicfiles.com/moco/moco_checkpoints/moco_v2_800ep/moco_v2_800ep_pretrain.pth.tar}}.
The publicly available pre-trained ResNet-50 encoder was trained using the ImageNet dataset under the MoCo-v2 framework.
Thus, we collect 10k random training samples from ImageNet and 10k samples from CIFAR-100 (serving as the non-member samples of the model).
We query the pre-trained ResNet-50 encoder and thus can obtain their feature vectors.
Subsequently, we calculated four different $p$-norms of the features.
As shown in Figure~\ref{fig:official_v2_hist}, we observe that there is a significant difference in the feature vector magnitude distributions between the member and non-member samples.
In addition, the $p$-norms value distributions of member and non-member samples can be approximated as two independent Gaussian distributions.
This behavior aligns closely with the differences highlighted in Figure~\ref{fig:local_traing_process}.

These observations in Figure~\ref{fig:local_traing_process} and Figure~\ref{fig:official_v2_hist} motivate us to construct a simple but effective membership inference attack method, named Embedding Lp-Norm Likelihood Attack (LpLA).

\subsection{Design of the Attack}
In this section, we consider an adversary who holds a partial ground-truth member dataset $\mathcal{T}_a \subset \mathcal{T}_p$ and can only obtain the output $v$ of target encoder $\mathcal{E}$ through black-box API.
In default, we set $p=2$ by default, while showing the effectiveness of the attack using other norms in Section 4.5.

\noindent \textbf{Feature extraction (Stage 1).} Firstly, for each sample (partial ground-truth member) held by the adversary $x \in \mathcal{T}_a$, the corresponding output feature vectors can be collected through black-box queries to the target encoder.
These feature vectors will be used in Stage 2 to estimate a $p$-norm distribution for the member class.
On the other hand, to construct a non-member dataset $\mathcal{T}_{an}$, the adversary can simply use random generation techniques to create samples matching the input format of the target model, such as generating random pixel images.
These samples are likewise queried through $\mathcal{E}$ in a black-box manner and subsequently collected for estimating a distribution of the non-member class.

\noindent \textbf{Likelihood estimation (Stage 2).} Based on the previous findings, we assume that the $p$-norms of the feature vectors for all member and non-member samples follow two independent Gaussian distributions.
To perform a membership inference attack on the victim sample under this assumption, the adversary needs to estimate these two normal distributions using a sufficient number of random samples collected like Stage1:
\begin{equation}
	\left\{
	\begin{aligned}
		L(x) &= \|\mathbf{v_x}\|_p = \sqrt[p]{v_1^p + v_2^p + \cdots + v_n^p}, \\
		L_{\text{m} } &\sim \mathcal{N}\left(\mu_{\text{member} }, \sigma_{\text{member} }^2\right), \\
		L_{\text{nm} } &\sim \mathcal{N}\left(\mu_{\text{non-member} }, \sigma_{\text{non-member} }^2\right),
	\end{aligned}
	\right.
\end{equation}
where $\mathbf{v_x}$ denotes a $n$ dimension feature vector output from the encoder with input $x$.
And $\mu_{\text{member}}$, $\sigma_{\text{member}}$ represent the mean and standard deviation of the $p$-norm values of member samples, respectively, $\mu_{\text{non-member}}$ and $\sigma_{\text{non-member}}$ are the corresponding parameters for non-member samples (To simplify notation, we adjusted some subscripts in this section: $\mu_{\text{m}}$ represents $\mu_{\text{member}}$, and $\mu_{\text{nm}}$ represents $\mu_{\text{non-member}}$ etc.).

Then, The adversary can estimate these parameters of two distributions based on the available $p$-norm values of member and non-member samples:
\begin{equation}
	\left\{
	\begin{aligned}
		\hat{\mu}_\text{m} &= \frac{1}{k} \sum_{i=1}^k L(x_{i}),\text{ } \hat{\sigma}_\text{m}^2 = \frac{1}{k-1} \sum_{i=1}^k \left(L(x_{i}) - \hat{\mu}_\text{m}\right)^2, \\
		\hat{\mu}_\text{nm} &= \frac{1}{q} \sum_{j=1}^q L(x_{j}),\text{ } \hat{\sigma}_\text{nm}^2 = \frac{1}{q-1} \sum_{j=1}^q \left(L(x_{j}) - \hat{\mu}_\text{nm}\right)^2, \\
		x_i &\in \mathcal{T}_a, \text{ } x_j \in \mathcal{T}_{an}.  \label{eq:mean-calculate}
	\end{aligned}
	\right.
\end{equation}

\noindent \textbf{Inferring membership (Stage 3).} Finally, for a given victim sample $x$ with $p$-norm value $L(x)$, the adversary can calculate a posterior probability of membership using Bayes' theorem:
\begin{equation}
	\left\{
	\begin{aligned}
		\mathcal{L}_{\text {m}}(x)&=\frac{1}{\sqrt{2 \pi \sigma_{\text {m }}^2}} \exp \left(-\frac{\left(L(x)-\mu_{\text {m }}\right)^2}{2 \sigma_{\text {m }}^2}\right), \\
		\mathcal{L}_{\text {nm}}(x)&=\frac{1}{\sqrt{2 \pi \sigma_{\text {nm }}^2}} \exp \left(-\frac{\left(L(x)-\mu_{\text {nm }}\right)^2}{2 \sigma_{\text {nm }}^2}\right), \\
		P(\text {m } \mid x)&=\frac{\mathcal{L}_{\text {m}}(x) P(\text {m})}{\mathcal{L}_{\text {m}}(x) P(\text {m})+\mathcal{L}_{\text {nm}}(x) P(\text {nm})}.
	\end{aligned}
	\right.
\end{equation}
Assuming there are equal prior probabilities $P(\text{m})=P(\text{nm})=0.5$, the decision criterion simplifies to:
\begin{equation}
	P(\operatorname{m} \mid x)>0.5 \quad \Longleftrightarrow \quad \mathcal{L}_{\text {m}}(x)>\mathcal{L}_{\text {nm}}(x).
\end{equation}

\takeawayN{\textbf{3:} The magnitude of the feature vector from members and non-members in encoders can be significantly different, which can be leveraged for membership inference.}

\subsection{Experiment Results of the Proposed Attack}
We now address \textbf{RQ4}: ``What are the advantages of the proposed attack compared to existing methods?''
To answer this question, we first compare the effectiveness of our proposed attack against the baseline attacks of Encoder-MI, SD-MI, and Fe-MI to show the effectiveness of our attack while with the lightweight advantage.
Then, we compare our attack with baseline attacks under the practical setting where the adversary has a partial training dataset and only black-box API query access to the target encoder model.
This aims to show the applicability advantage of our attack over the baseline attacks.

\noindent \textbf{Attack effectiveness.} We first show the MI attack performance against encoders in Table \ref{tab:Result_datasets}, which is conducted on the ResNet-50 backbone across three different datasets and MoCo frameworks.
As we can see, LpLA demonstrates sufficiently competitive attack performance across various dataset experiments.
Unfortunately, LpLA, like existing methods, is unable to effectively infer membership status in the MoCo-v1 stage.
However, under the MoCo-v2 framework, LpLA starts to show attack performance on par with classic feature-based and similarity-based methods.
As the contrastive learning framework is further updated, we can observe that LpLA achieves the best attack performance on both CIFAR-100 and Tiny-Imagenet.
\begin{table}[t]
	\centering
	\caption{Attack accuracy (TPR@0.1\%FPR) against encoders.}
	\resizebox{0.5\textwidth}{!}{
		\begin{tabular}{cccccc}
			\toprule
			\textbf{Dataset} & \textbf{Model} & \textbf{Encoder-MI} & \textbf{SD-MI} & \textbf{Fe-MI} & \textbf{LpLA(ours)} \\
			\midrule
			\multirow{3}[2]{*}{\makecell{CIFAR-10}} & MoCo-v1 & 0.494(0.098) & 0.501(0.101) & 0.497(0.099) & 0.496(0.099)  \\
			& MoCo-v2 & 0.607(0.146) & 0.718(0.253) & 0.692(0.201) & \textbf{0.723}(\textbf{0.259})  \\
			& MoCo-v3 & \textbf{0.960}(\textbf{3.181}) & 0.914(1.045) & 0.910(1.075) & 0.907(0.790)  \\
			\midrule
			\multirow{3}[2]{*}{\makecell{CIFAR-100}} & MoCo-v1 & 0.512(0.104) & 0.507(0.105) & 0.506(0.102) & 0.496(0.098)  \\
			& MoCo-v2 & 0.627(0.173) & \textbf{0.783}(0.326) & 0.750(0.318) & 0.765(\textbf{0.415})  \\
			& MoCo-v3 & 0.946(0.919) & 0.915(\textbf{2.377}) & 0.953(2.216) & \textbf{0.968}(1.795)  \\
			\midrule
			\multirow{3}[2]{*}{\makecell{Tiny-Imagenet}} & MoCo-v1 & 0.507(0.103) & 0.496(0.099) & 0.498(0.099) & 0.501(0.103)  \\
			& MoCo-v2 & 0.583(0.142) & \textbf{0.777}(0.384) & 0.754(0.347) & 0.769(\textbf{0.436})  \\
			& MoCo-v3 & 0.965(3.368) & 0.979(\textbf{5.858}) & 0.976(4.285) & \textbf{0.982}(3.279)  \\
			\bottomrule
	\end{tabular}}
	\label{tab:Result_datasets}
\end{table}

Additionally, an interesting phenomenon is observed that in several attack scenarios where LpLA does not achieve the best performance, SD-MI shows higher accuracy than other attack methods under the MoCo-v2 framework, and it's later surpassed by LpLA when the framework is updated to MoCo-v3.
During the update, there is indeed a certain characteristic of the output features that was more straightforwardly captured by LpLA, thus bridging the gap between LpLA and similarity-based attacks.

We also implement our LpLA against the official pre-trained encoder for validation.
Due to different training settings, we did not compare the experimental results with those based on our locally trained target models.
Table~\ref{tab:Result_LpLA_on_official} shows that LpLA also achieves strong attack performance against the official pre-trained encoder.
\begin{table}[t]
	\centering
	\caption{LpLA against official pre-trained ResNet-50.}
	\resizebox{0.45\textwidth}{!}{
		\begin{tabular}{cccccc}
			\toprule
			\textbf{Model} & \textbf{Epochs} & \textbf{p=0} & \textbf{p=1} & \textbf{p=2} & \textbf{p=3} \\
			\midrule
			MoCo-v2(official) & 800  & 0.732 & \textbf{0.779} & 0.688 & 0.681  \\
			MoCo-v3(official) & 300 & 0.535 & 0.539 & 0.609 & \textbf{0.674}  \\
			\bottomrule
	\end{tabular}}
	\label{tab:Result_LpLA_on_official}
\end{table}

\noindent \textbf{Attack applicability.} Compared to existing attack approaches, LpLA does not require training neural networks as the binary attack classifier.
Consequently, LpLA demands far fewer samples for constructing an attack model compared to neural network-based methods (e.g., EncoderMI, SD-MI, and Fe-MI).
This is because LpLA directly estimates the likelihood of $p$-norm of feature vectors, leveraging statistical methods to infer differences in data distributions.
Moreover, LpLA is computationally efficient.
Traditional neural network-based attacks often involve higher resource demands for optimization, sometimes requiring additional queries and similarity calculations, resulting in significant computational costs.
In contrast, LpLA only involves $p$-norm computation and distribution estimation after queries, making it lightweight and computationally efficient.
The fewer attack requirements make LpLA more applicable in different attack scenarios than the existing baseline attacks. 

We qualitatively compare our attack with existing baseline attacks in Table~\ref{tab:comparison}.
As we can see, LpLA reduces both sample and computational requirements, lowering the technical barrier for the adversary while broadening the applicability of membership inference attacks.
We also quantitatively evaluate the attack performance of the scenarios where the number of member samples and model query access is limited, as follows.

\begin{table*}[tbp]
	\caption{Qualitatively comparison between our attack with existing baseline attacks.}
	\label{tab:comparison}
	\begin{tabular}{ccccc}
		\toprule
		\textbf{Attacker}  &\textbf{Ground-truth samples}   &\textbf{Query volumes}    &\textbf{Computational requirement}   &\textbf{Other}\\
		\midrule
		Encoder-MI      & $2*2k$   & $2*10*2k$           & Similarity \& MLP-training          & Data augmentation\\
		SD-MI           & $2*2k$   & $2*2k+2k(Anchors)$  & Similarity \& Two-MLP-training      & Auxiliary dataset\\
		FE-MI           & $2*2k$   & $2*2k$              & MLP-training                        & (None)\\
		LpLA(Ours)      & $2*2k$   & $2*2k$              & L$p$-norm \& Likelihood-estimation  & (None)\\
		\bottomrule
	\end{tabular}
\end{table*}

\noindent $\bullet$ \textbf{Attack with limited member samples.} In real-world scenarios, an adversary often faces various limitations due to insufficient information about the target model, such as the inability to obtain enough ground-truth member samples, which can significantly weaken the attack's effectiveness.
Furthermore, in the context of MIA against encoder models, because the size of unlabeled datasets used for self-supervised training is often enormous, it's difficult to collect enough shadow datasets that are identically distributed to the training set to train shadow models.
Therefore, the number of ground-truth member samples that an adversary can obtain becomes a significant challenge.
\begin{figure}[t]
	\centering
	\subfigure[ResNet-50 + MoCo-v2]{\includegraphics[width=4.0cm]{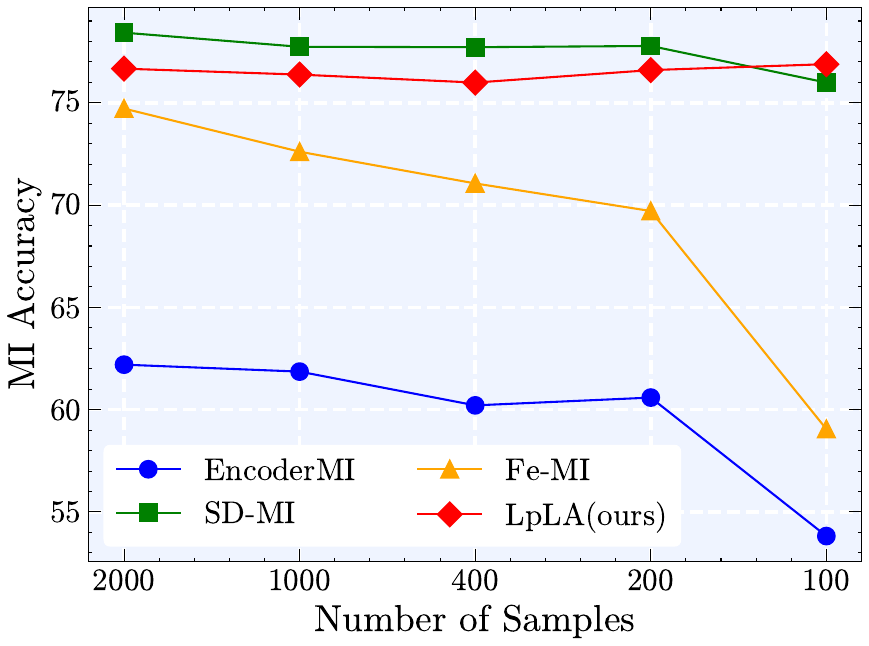}}
	\subfigure[ResNet-50 + MoCo-v3]{\includegraphics[width=4.0cm]{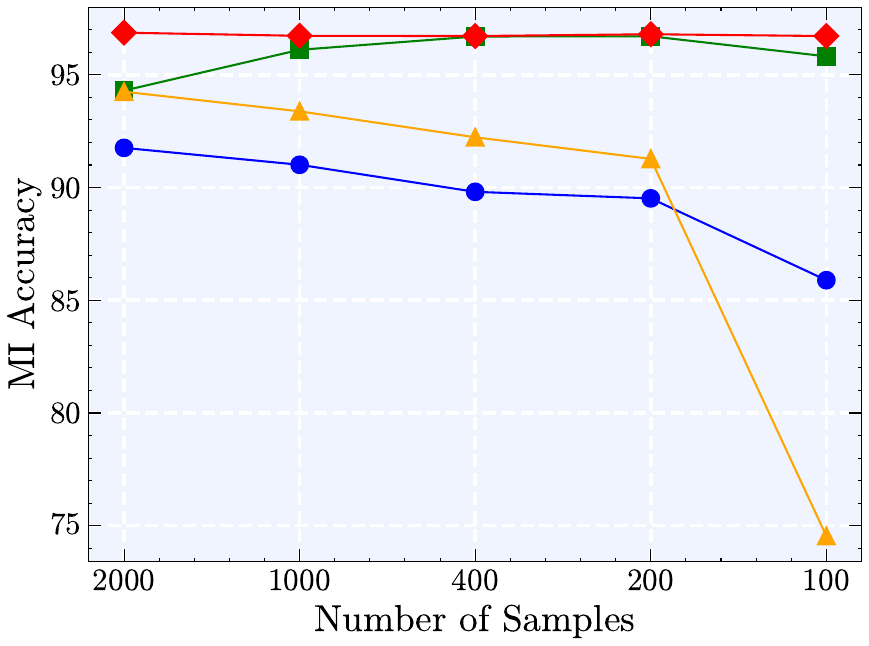}}
	\vspace{-0.2cm}
	\caption{An illustration of the attack performance with limited attack dataset. There is a significant decreasing trend of EncoderMI and Fe-MI when gradually reducing the scale of the attack dataset, while SD-MI and LpLA perform relatively more robustly.}
	\label{fig:Result_numbers}
\end{figure}

Figure~\ref{fig:Result_numbers} shows the attack performance achieved by different attacks with different numbers of ground-truth member samples.
The experimental results indicate that among all attack methods, LpLA demonstrates sufficiently competitive MIA performance (in most cases being the best attack method, and in a few cases being the second-best method), showing the superiority of LpLA in attack effectiveness.
Additionally, the experimental results also show that when the size of the ground-truth member set held by the adversary decreases, the attack effectiveness also tends to decline.
However, among these four methods, LpLA exhibits the most robust attack performance, as even when the data proportion is significantly reduced, the degree of decline in LpLA's attack effectiveness is still the smallest.

\noindent $\bullet$  \textbf{Attack with limited query volumes.} Furthermore, we also consider that in real-world scenarios, an adversary often has to bear expensive query costs when accessing black-box models, or a privacy-aware victim may limit the number of API queries.
Therefore, we further consider the attack performance that each method can achieve under different query scale settings when the adversary faces a limited number of queries to the target model.
\begin{figure}[t]
	\centering
	\subfigure[ResNet-50 + MoCo-v2]{\includegraphics[width=4.0cm]{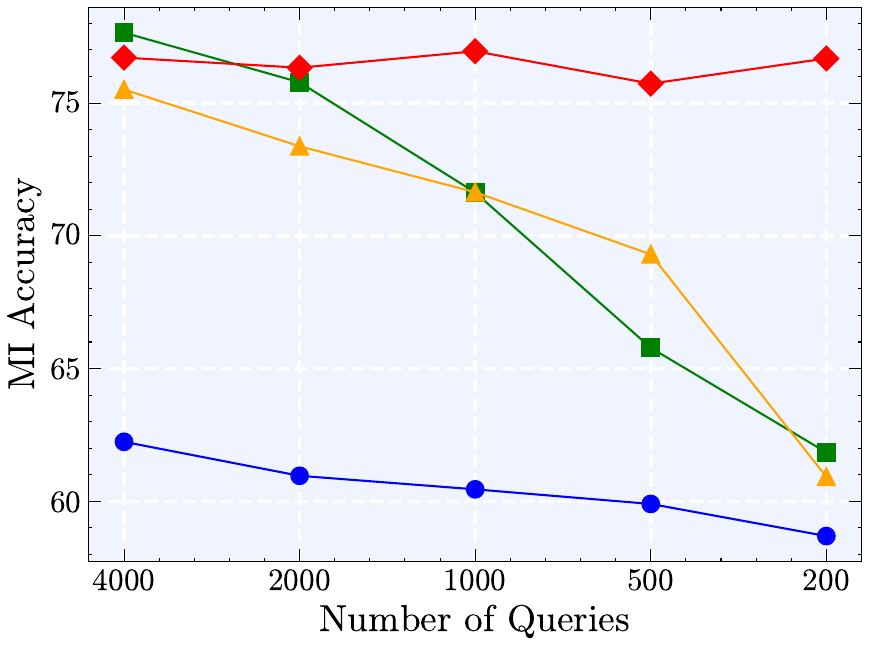}}
	\subfigure[ResNet-50 + MoCo-v3]{\includegraphics[width=4.0cm]{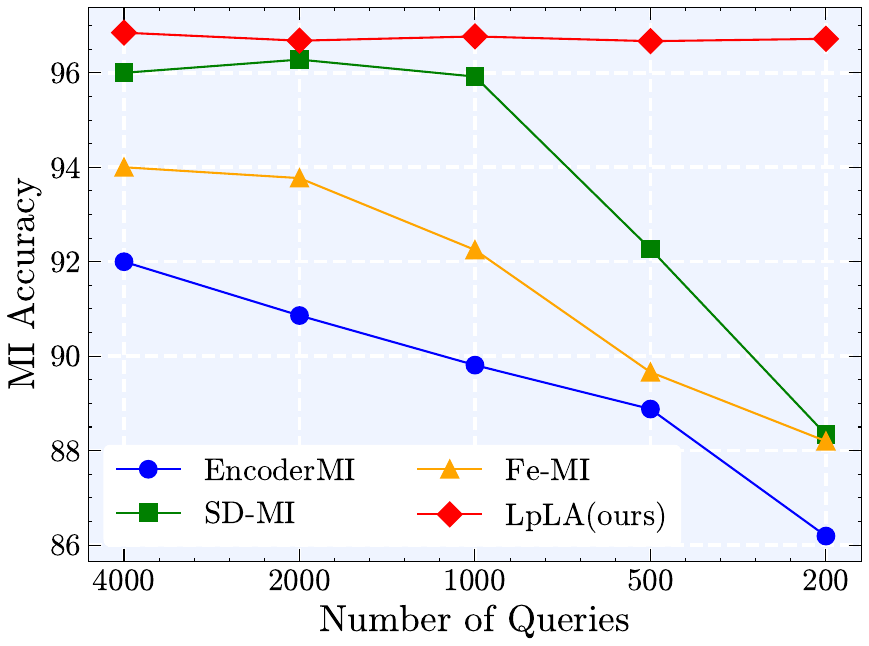}}
	\vspace{-0.2cm}
	\caption{An illustration of the attack performance with limited API query volumes. There is only LpLA performs robust and efficiently under gradually limited API query volumes, while other attacks perform insufficiently.}
	\label{fig:Result_queries}
\end{figure}

Figure~\ref{fig:Result_queries} shows the attack performance achieved when four different attack methods are used with a limited number of queries.
It can be observed that when the API query numbers are severely limited, existing methods exhibit a more severe decline trend compared to the cases when the number of ground-truth member samples is limited.
It's because when using multiple data augmentation samples or anchors to construct similarity-based attack features, more API query costs need to be paid, leading existing methods to be more easily affected by the constrained conditions in this scenario.

On the other hand, LpLA can still achieve effective attack performance, under conditions where the adversary's knowledge is highly limited.
This is because when LpLA conducts inference attacks, it's based on likelihood estimation, and mainly relies on the differences in data statistical distribution.
Thus, LpLA estimating the distributions with likelihood does not require as many samples as training a neural network, which allows it to demonstrate stronger adaptability in knowledge-limited scenarios.

\takeawayN{\textbf{4:} Being lightweight, the proposed attack, LpLA, has comparable attack performance against existing baseline attacks.
	In scenarios where only a limited number of member samples and query volumes are available, LpLA outperforms existing baseline attacks on encoder models.}

\subsection{Ablation}
We further conduct ablation study experiments to evaluate how the choice of $p$-norm and distribution likelihood can affect the effectiveness of the proposed attack.

\begin{figure}[tbp]
	\centering
	\includegraphics[width=0.37\textwidth]{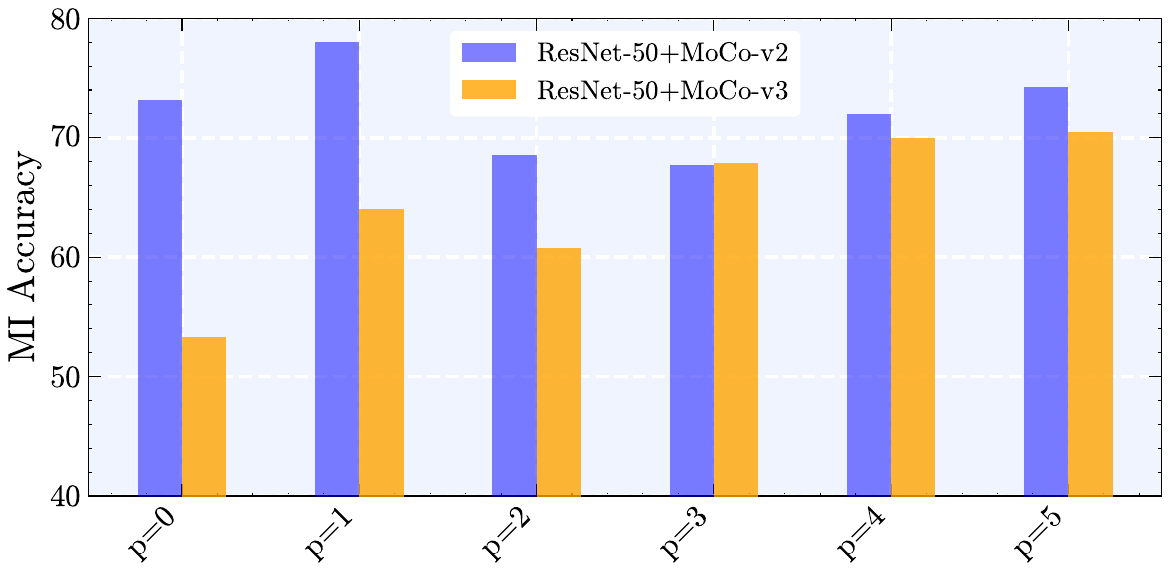}
	\caption{\textbf{An illustration of the performance of LpLA against ResNet-50 + MoCo-v2\&3 with different $p$. There is  an insufficient performance when $p=0$, and shows similarly satisfying performance with other $p$.}}
	\label{fig:Pvalue_target}
\end{figure}

\noindent \textbf{$p$-value selection.} Figure~\ref{fig:Pvalue_target} shows the results of LpLA on ResNet-50 pre-trained in MoCo-v2 and MoCo-v3 frameworks with various $p$-values.
By comparing the attack performance across norms from p=0 to 5, we can see that using the L0-norm of feature vectors yields highly suboptimal attack results.
However, as the $p$-value increases, the attack accuracy improves correspondingly.
When the $p$-value reaches 2, the attack accuracy saturates, showing no significant improvement with further increases in $p$.

\begin{figure}[t]
	\centering
	\subfigure[MI against MoCo-v2]{\includegraphics[width=4.0cm]{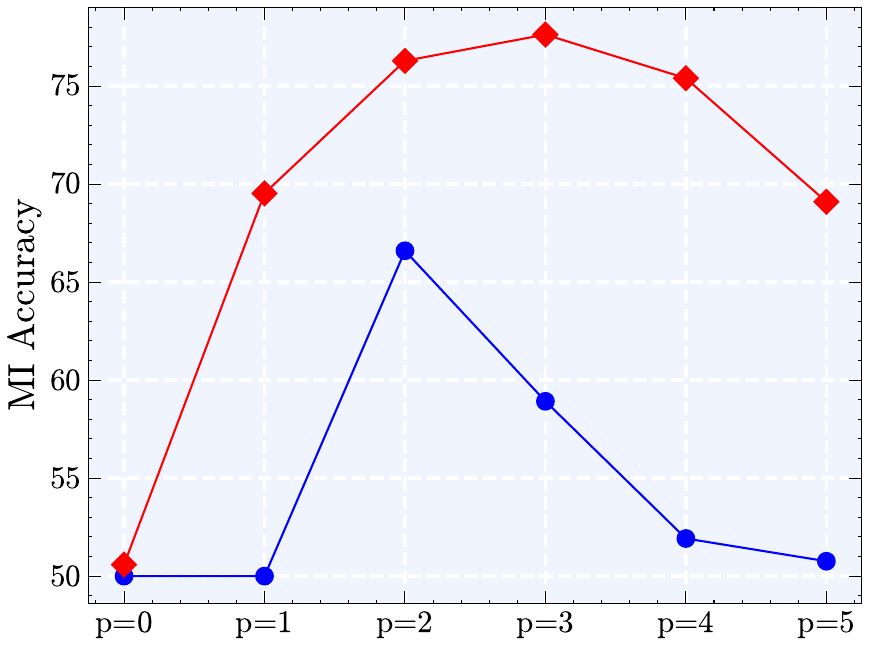}}
	\subfigure[MI against MoCo-v3]{\includegraphics[width=4.0cm]{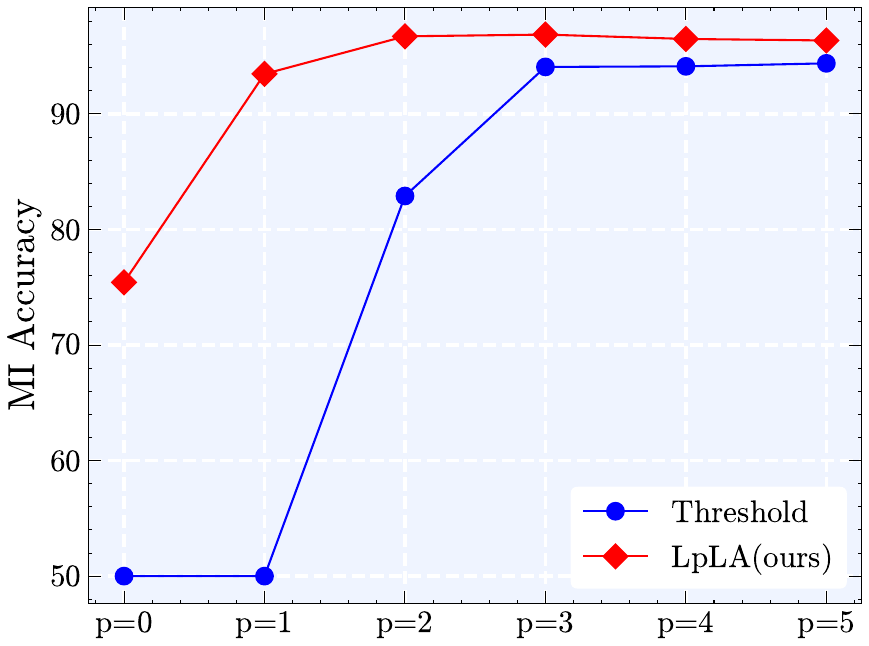}}
	\vspace{-0.2cm}
	\caption{An illustration of the comparison between using $p$-norm for estimation and threshold-based MI against ResNet-50 + MoCo-v2\&3. LpLA shows a significant advantage over threshold-based attack, because LpLA uses not only the information of $p$-norm's mean, but uses estimated variance which helps attack more accurately and robustly.}
	\label{fig:Result_threshold}
\end{figure}

\noindent \textbf{Distribution likelihood estimation.} Figure~\ref{fig:Result_threshold} shows the results when simply use $p$-value as a threshold to perform MIA against encoder pre-trained in MoCo-v2 and MoCo-v3 frameworks with various $p$-values.
We perform a simple threshold-based MI here, where the mean $p$-value of member and non-member is calculated like equation \eqref{eq:mean-calculate}.
Comparing LpLA and naive threshold-based MI, it can be observed that likelihood estimation helps capture the information of attack signals more accurately and robustly.

\subsection{Limitations}
Despite LpLA demonstrating superior attack performance and robustness across diverse datasets and model architectures, several limitations warrant consideration:
(1) The resilience of LpLA against existing defense mechanisms (e.g., differential privacy or adversarial regularization) remains unexamined, leaving its robustness under countermeasures uncertain;
(2) Experiments are confined to image datasets (e.g., CIFAR, Tiny-ImageNet), with no validation on text or audio modalities, limiting insights into cross-domain applicability.
Future work will assess defenses' impacts on SSL's performance-privacy trade-off and expand research to text/audio modalities to ensure generalizable analysis.

\section{Conclusion}
We delved deeply into the issue of membership privacy in this work, especially focusing on the issue of membership inference attacks against encoder models within contrastive learning.
It comprehensively reveals the impact of different model architectures on the leakage of membership information through both theoretical analysis and experimental validation.
Based on this, we propose a membership inference attack method named LpLA (Embedding Lp-Norm Likelihood Attack), which is motivated by the different distributions of member and non-member's embedding $p$-norm values.
Experimental results indicate that more complex model architectures, while enhancing the feature extraction capabilities of encoder models, also exacerbate the risk of membership privacy leakage.
Furthermore, this research demonstrates the effectiveness of using the $p$-norm of feature vectors in MI attacks. LpLA not only matches or even surpasses existing methods in performance, but most importantly, LpLA exhibits greater robustness, especially under conditions where the adversary's knowledge is severely limited.

This research not only broadens the scope of investigations into membership privacy issues in contrastive learning but also lays a solid foundation for future studies on privacy protection.
It is hoped that this work will inspire more attention to the privacy risks associated with self-supervised learning models, such as performing privacy attacks and implementing protections across a wider range of deep learning models.
%, it could be based on some specific information dug from output using statistical theories.

\begin{acks}
    Haizhuan's work was supported in part by National Natural Science Foundation of China (No. 12271464), Hunan Provincial Natural Science Foundation of China (No. 2023JJ10038), the Innovative Research Group Project of Natural Science Foundation of Hunan Province of China (No. 2024JJ1008). We also gratefully acknowledge the support of the High-Performance Computing Platform of Xiangtan University for providing computational resources used in this work.
\end{acks}

\bibliographystyle{ACM-Reference-Format}
%\bibliography{sample-base}
\bibliography{ref}

\appendix



%Lorem ipsum dolor sit amet, consectetur adipiscing elit. Morbi


%Etiam commodo feugiat nisl pulvinar pellentesque. Etiam auctor sodales

\end{document}